\documentclass[11pt]{article}

\usepackage[final]{acl}

\usepackage{times}
\usepackage{latexsym}
\usepackage{amsmath} 
\usepackage{amssymb} 
\usepackage{algorithmicx,algorithm}

\usepackage[T1]{fontenc}

\usepackage[utf8]{inputenc}

\usepackage{microtype}

\usepackage{inconsolata}

\usepackage{graphicx}
\usepackage{xspace}
\usepackage{makecell}
\usepackage{array}
\usepackage{colortbl}
\usepackage{multirow}
\usepackage{subcaption}
\usepackage{tabularx}
\usepackage[dvipsnames]{xcolor} 
\usepackage{color}
\usepackage{pifont}
\usepackage[utf8]{inputenc}
\usepackage[T1]{fontenc}
\usepackage{CJKutf8}
\usepackage{enumitem}
\usepackage{diagbox}
\usepackage{listings}
\usepackage[flushleft]{threeparttable}
\usepackage{booktabs}
\usepackage{wrapfig}
\usepackage{marvosym}

\newcommand{\edit}[1]{\textcolor{black}{#1}}

\newcommand{\ourmethod}{{\fontfamily{lmtt}\selectfont \textbf{ST-EVO}}\xspace}
\newcommand{\llmname}[1]{{\texttt{#1}}}

\definecolor{fbApp}{HTML}{ffe4e3}
\definecolor{mydarkblue}{rgb}{0,0.3,0.9}

\newcommand{\rowcr}{\rowcolor{fbApp}}

\newcommand{\rowcb}{\rowcolor{CadetBlue!10}} 
\newcommand{\rowcg}{\rowcolor{gray!10}}

\newcommand{\first}[1]{\textcolor{red}{\textbf{#1}}}
\newcommand{\second}[1]{\textcolor{blue}{\underline{#1}}}

\newcommand{\blue}[1]{$_{\color{BlueGreen}\downarrow #1}$}
\newcommand{\red}[1]{$_{\color{RedOrange}\uparrow #1}$}

\title{ST-EVO: Towards Generative Spatio-Temporal Evolution of \\ Multi-Agent Communication Topologies}

\author{
 \textbf{Xingjian Wu}\textsuperscript{*},
 \textbf{Xvyuan Liu}\textsuperscript{*},
 \textbf{Junkai Lu}\textsuperscript{*},
 \textbf{Siyuan Wang},
 \textbf{Xiangfei Qiu}, \\
 \textbf{Yang Shu},
 \textbf{Jilin Hu},
 \textbf{Chenjuan Guo},
 \textbf{Bin Yang}\textsuperscript{\Letter}
\\
 East China Normal University
\\
   \{xjwu, xvyuanliu, jklu, sywang, xfqiu\}@stu.ecnu.edu.cn, \\ 
   \{yshu, jlhu, cjguo, byang\}@dase.ecnu.edu.cn
 }

\begin{document}
\maketitle
\begin{abstract}
LLM-powered Multi-Agent Systems (MAS) have emerged as an effective approach towards collaborative intelligence, and have attracted wide research interests. Among them, ``self-evolving'' MAS, treated as a more flexible and powerful technical route, can construct task-adaptive workflows or communication topologies, instead of relying on a predefined static structure template. Current self-evolving MAS mainly focus on Spatial Evolving or Temporal Evolving paradigm, which only considers the single dimension of evolution and does not fully incentivize LLMs' collaborative capability. In this work, we start from a novel Spatio-Temporal perspective by proposing \ourmethod, which supports dialogue-wise communication scheduling with a compact yet powerful flow-matching based Scheduler. To make precise Spatio-Temporal scheduling, the Scheduler can also perceive the uncertainty of MAS, and possesses self-feedback ability to learn from accumulated experience. Extensive experiments on nine benchmarks demonstrate the state-of-the-art performance of \ourmethod, achieving about 5\%--25\% accuracy improvement.
\end{abstract}

\section{Introduction}
LLM-powered agentic systems~\citep{autogpt,langgraph,wu2024autogen}, which integrate LLMs' reasoning capabilities with external functionalities, have exhibited impressive performance and facilitated a surge of practical applications like information search~\citep{li2025search}, data analysis~\citep{TimeART}, code generation~\citep{zheng2023codegeex}, deep research~\citep{zheng2025deepresearcher} and autonomous driving~\citep{jin2023surrealdriver}. To further improve the performance and robustness in complex scenarios, multi-agent systems (MAS) are studied to achieve collaborative intelligence. Generally, multi-agent systems~\citep{ding2024large,han2024llm} are customized for a certain type of task, with a meticulously-designed workflow or communication topology, focusing on how agents effectively utilize, exchange, and induce information reciprocally.
\begin{figure}[t]
    \centering
\includegraphics[width=1\linewidth]{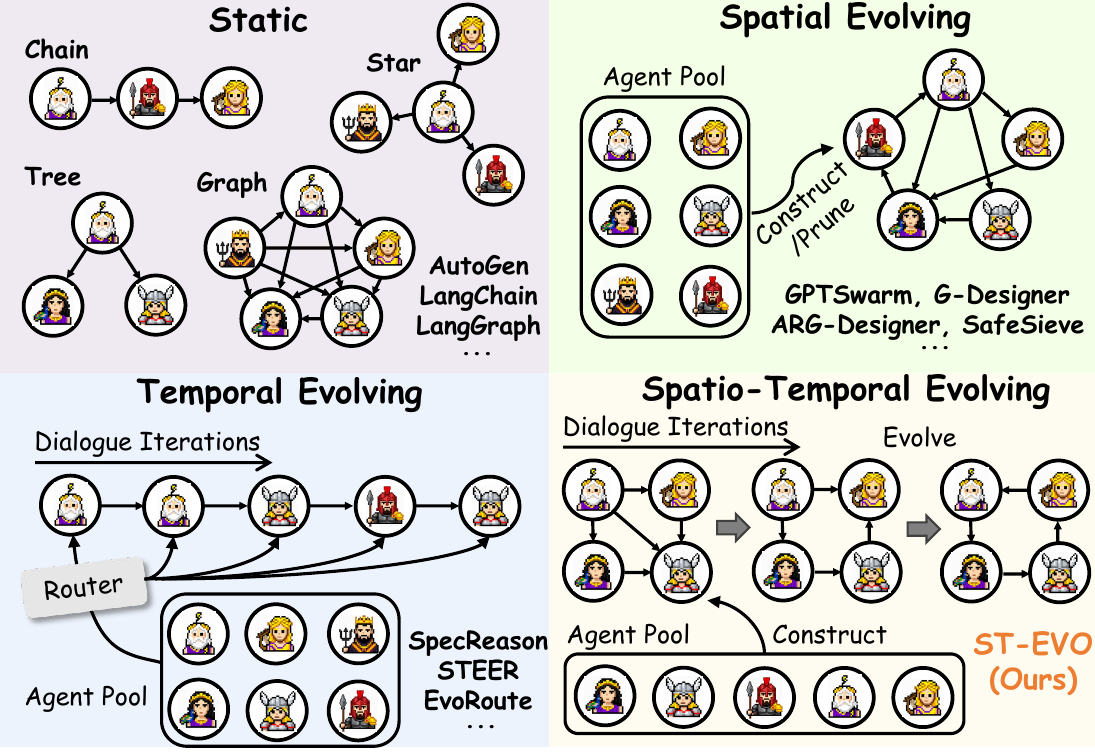}
    \caption{Existing multi-agent systems. Our proposed \ourmethod pioneers the exploration of Spatio-Temporal Evolving multi-agent systems.}
\label{fig: intro}
\end{figure}

Classic multi-agent systems often adopt static topologies--see Figure~\ref{fig: intro} (top left), which organizes agents as sequential workflows (chain)~\citep{wei2022chain,hong2023metagpt}, task splitting mode (tree)~\citep{yao2023tree,yao2022react}, or task assignment mode (star)~\citep{wu2024autogen}. Since they are all special cases of graphs, some recent multi-agent systems~\citep{langgraph,wu2024autogen} adopt graphs as their template structures and demonstrate remarkable performance. However, these task-specific, static designs rely on expert intervention, and lack automation and adaptation when scenario changes. Therefore, current research interests gradually shift towards dynamic workflow orchestration~\citep{niu2025flow,zhangaflow,lee2025confidence} and communication topology design~\citep{GDesigner,li2025assemble,zhang2025safesieve}, aiming at task-specific or query-specific automated fine-grained customization, known as ``\textit{self-evolving}'' MAS~\citep{fang2025comprehensive,wang2025evoagentx}.

As shown in Figure~\ref{fig: intro}, we categorize existing self-evolving multi-agent systems into ``Spatial Evolving'' and ``Temporal Evolving''. Considering the Spatial Evolving MAS, the mainstream approach is to first formulate the communication topology as a graph, then learn from some cases to train a ``designer'' (often compact neural networks), which can design topologies for specific tasks~\citep{liu2023dynamic,zhuge2024gptswarm} or queries~\citep{GDesigner,li2025assemble,zhang2025safesieve}. However, the designed topologies are fixed throughout the entire conversation process for a query. That means, if a complex query needs multiple dialogue iterations to solve, the communication topology cannot be dynamically tuned during iterations to allocate computing resources and functions among different agent nodes, \textit{which leads to low flexibility and robustness, and may cause high token consumption.} On the other hand, though the Temporal Evolving MAS makes temporal split on workflows, most of them focus on routing single agent in the proper position~\citep{niu2025flow,zhangaflow,lee2025confidence}, \textit{which neglects the potential of multi-agent collaboration in key dialogue iterations.} 

The questions in single-dimension evolution posed above can be explicitly summarized as: \ding{182} \textbf{Accuracy.} How to fully unleash the collaborative capabilities of MAS in different dialogue iterations to fulfill the changing demand? \ding{183} \textbf{Stability.} How to ensure the scheduling is reasonable and robust? \ding{184} \textbf{Efficiency.} How to ensure efficiency considering the token consumption of MAS and the cost of scheduling? To tackle these problems, we propose \ourmethod, the first MAS with the capability of Spatio-Temporal Evolution, achieving dialogue-wise temporal scheduling for spatial communication topologies. 

The motivation of \ourmethod is to precisely schedule multi-agent collaborations from both temporal and spatial dimensions, preserving accuracy, stability, and efficiency. To achieve this, we believe the core elements lie in: \ding{168} \textbf{Appropriate data structures.} The structure should be able to well maintain the Spatio-Temporal trajectories, providing the basis for scheduling. \ding{171} \textbf{A compact but effective scheduler.} The scheduling is frequent, and the demand for accuracy is high, thus calling for a lightweight and powerful scheduler. \ding{169} \textbf{System state perception.} The scheduler should consider the system state, such as whether it is reliable or not, thus leading to correct scheduling scheme. \ding{170} \textbf{Self-feedback ability.} The system needs to accumulate experience and learn underlying scheduling principles from it.

Based on above-mentioned design principles, \ourmethod adopts \textit{Spatio-Temporal Graph} to maintain the scheduling trajectories, which is an important structure in Spatio-Temporal data mining~\citep{wang2020deep,yang2024survey}, able to characterize the spatial correlations (agent collaboration) at each temporal index (dialogue iteration). Considering the scheduler, we meticulously devise a compact neural network, driven by an MLP-based \textit{Flow-Matching}~\citep{FlowMatch} module, which possesses strong generative mechanism while keeping lightweight, and can work as a stochastic interpolant~\citep{wu2026aurora} to directly evolve the communication topology temporally.

Additionally, to perceive the system state and provide self-feedback ability, we devise mechanisms driven by both entropy and experience, then internalize them through preference optimization. First, we supervise the system state with the help of \textit{Entropy Distribution} to evaluate the uncertainty of the MAS, and we internalize it into the scheduler via reinforcement learning. Second, considering the self-feedback ability, we draw inspiration from \textit{Experience Learning}~\citep{zhang2025agent,tang2025chemagent}, and construct a retrieve-augment module to accumulate successful scheduling trajectories as experience. During training, we also internalize the experience, which retrieves corresponding historical scheduling trajectories as reference, and regularizes the current scheduling scheme.

Our Contributions can be summarized as:

\begin{itemize}[leftmargin=2em,itemsep=-0.1em]
\item[\ding{182}] We introduce \ourmethod, pioneering the exploration of Spatio-Temporal Evolving MAS. It can support dialogue-wise real-time scheduling through generative Spatio-Temporal evolution of communication topologies.
\item[\ding{183}] We devise a compact generative scheduler driven by Flow-Matching. It can learn underlying scheduling principles from entropy distribution and historical experience, thus possessing strong query-aware and dialogue-wise scheduling capability. 
\item[\ding{184}] Comprehensive experiments on 9 benchmarks from multiple real-world tasks demonstrate the state-of-the-art performance of our proposed \ourmethod. All the code and datasets are available at the anonymous repository \href{https://github.com/decisionintelligence/ST-EVO}{https://github.com/decisionintelligence/ST-EVO}.
\end{itemize}

\section{Related Works}
\subsection{LLM-powered Multi-Agent Systems} 

Recent studies introduce a surge of multi-agent systems to incentivize LLMs' swarm intelligence in complex scenarios. The communication topology occupies a central position in them and includes multiple structures. Among them, non-interactive systems organize agents parallelly without explicit communication, such as Mixture-of-Agents~\citep{wangmixture} and LLM-Debate~\citep{estornell2024multi}; Chain MAS arranges agents sequentially as a workflow, such as MetaGPT~\citep{hong2023metagpt} and ChatDev~\citep{qian2024chatdev}; Tree MAS works in a task-splitting mode with a chief node and hierarchical child nodes~\citep{yao2023tree}; Star MAS works in a task-assignment mode with a central node and multiple edge nodes, seen in AutoGen~\citep{wu2024autogen} and MiniGrid~\citep{zhou2024large}; Graph MAS, as the most general form, can represent all of above-posed topologies and are widely applied~\citep{langgraph,zhuge2024gptswarm,wu2024autogen}.

\subsection{Self-evolving Multi-Agent Systems} 
To improve efficiency, recent research interests shift towards ``self-evolving'' MAS~\citep{wang2025evoagentx,fang2025comprehensive}, which can contribute to MAS customization for changing tasks. These studies characterize multi-agent communication topologies as graphs~\cite{liu2023dynamic,zhuge2024gptswarm,GDesigner} to organize and represent relationships between agents. Compared with conventional static communication topologies, they support task-aware~\citep{liu2023dynamic,zhuge2024gptswarm} or even query-aware~\citep{GDesigner,li2025assemble,zhang2025safesieve} dynamic communication topology customization. We categorize them into Spatial Evolving and Temporal Evolving MAS. For Spatial Evolving MAS, they mainly focus on the performance, robustness, efficiency, and redundancy~\citep{zhang2025safesieve} when customizing the topology. However, they pay little attention to temporal scheduling which is crucial in MAS with dialogue-wise varying resource demands. For Temporal Evolving MAS~\citep{zhangaflow,niu2025flow,lee2025confidence}, though they consider the temporal scheduling in different iterations, most of them neglect utilizing the collaborative capability of multiple agents inside each iteration. In this work, we propose \ourmethod, the first Spatio-Temporal Evolving MAS, which supports temporal scheduling for spatial communication topologies.

\section{Formalization}
\label{sec: formalization}
In this section, we formalize key concepts in \ourmethod from a spatio-temporal perspective, and introduce the notations for clear clarification.
\subsection{Communication Topology}
In \ourmethod, a multi-agent system is modeled as a Spatio-Temporal Graph $\mathcal{G} = \{\mathcal{G}_t\}_{t=1}^T=\{(\mathcal{V}, \mathcal{E}_t,\mathrm{A}_t)\}_{t=1}^T$, where $t$ records the dialogue iteration. Note that the total iterations $T$ is flexible and not pre-defined. $\mathcal{V} = \{v_1, v_2, \cdots, v_N\}$ denotes $N$ agent nodes in MAS:
\begin{gather}
    v_i = \{\text{LLM}_i, \text{Profile}_i, \text{Tool}_i, \text{State}_i \},\label{eq1}
\end{gather}
where each agent $v_i$ is instantiated through a LLM with Profile (system prompts) and Tool, and its historical memory is maintained in State. During operation, each agent $v_t$ treats $\text{Profile}_i$ and $\text{State}_i$ as a prefix, and receives user prompts or responses from other agents in MAS.

In \ourmethod, the communication topology $\mathcal{G}_t$ changes over time, contributing to the varying communication link $\mathcal{E}_t$, which is described via the adjacent matrix $\mathrm{A}_t\in \{0,1\}^{N\times N}$, where $\mathrm{A}_t[i,j]=1$ denotes the edge $e_t[i,j] \in \mathcal{E}_t$ is connected and the information flows from $v_i$ to $v_j$, vice versa.

\subsection{Scheduling Pipeline}

Given a query $\mathcal{Q}$, \ourmethod first constructs the communication topology $\mathcal{G}_1$ from an initial anchor topology $\mathcal{G}_0$ for the first dialogue iteration. For simple questions, they can be easily tackled at once. For harder ones, multi-iteration dialogues are executed and fine-grained scheduling is planned through a Scheduler.  

In the $t$-th dialogue, the Scheduler considers the problem $\mathcal{Q}$, the iteration $t$, and the topology from the last iteration $\mathcal{G}_{t-1}$ to determine the topology of this iteration $\mathcal{G}_{t}$:
\begin{gather}
    \mathcal{G}_{t} = \text{Scheduler}(\mathcal{Q}, t, \mathcal{G}_{t-1})
\end{gather}
Then the communication topology $\mathcal{G}_t$ is compiled into a \textit{topological} execution order $\mathcal{S}_t$ of the $N$ agents in MAS:
\begin{gather}
    \mathcal{S}_t = [v^t_{s_1}, v^t_{s_2}, \cdots, v^t_{s_N}],\\
    \text{s.t.} \ \forall j>i, \ v^t_{s_j} \notin \mathcal{N}_{\text{in}}(v^t_{s_{i}})
\end{gather}
After \ourmethod executes $T$ iterations, the final solution is formed or partially formed inside the agent states, an aggregation operation like majority voting is applied to summarize for the output.

\begin{figure*}[t]
    \centering
\includegraphics[width=1\linewidth]{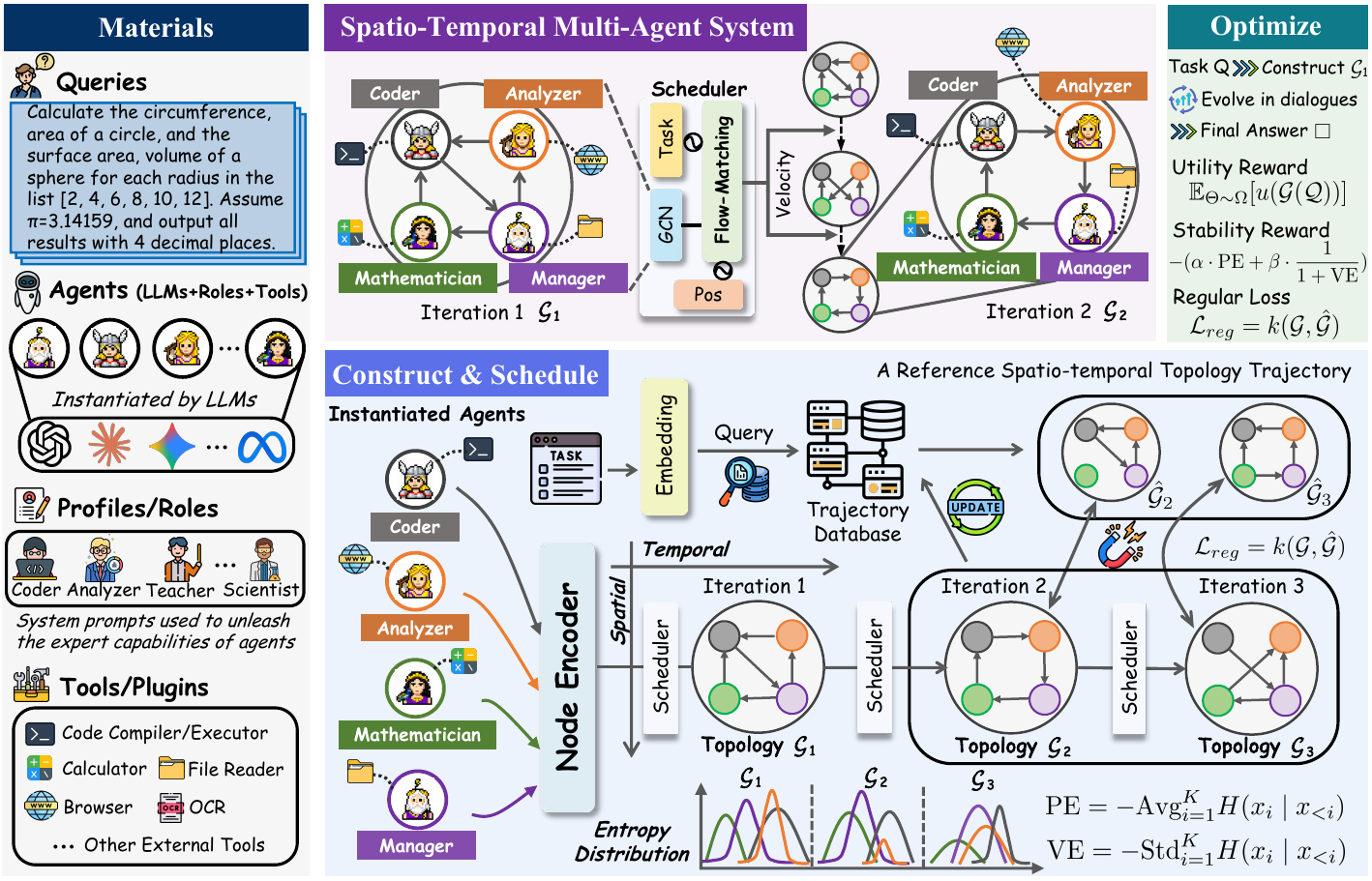}
    \caption{\edit{The overview of \ourmethod. The framework follows a macro-to-micro logic: Construct \& Schedule outlines the core macro-workflow where the Scheduler iteratively refines the communication topology $\mathcal{G}_i$. Supporting this, Materials details the foundational components used to instantiate agents, while the Spatio-Temporal Multi-Agent System illustrates the step-wise topology evolution via graph flow matching. Finally, Optimize formulates the trajectory-retrieval regularized loss and the stability reward that align the Scheduler's preferences. }}
\label{fig: overview}
\end{figure*}

\section{ST-EVO}
As illustrated in Figure~\ref{fig: overview}, our proposed \ourmethod builds multi-agent systems from a spatio-temporal perspective. (1) \ourmethod supports query-aware MAS construction, which builds agents with provided LLMs, profiles and Tools, and generates the initial communication topology based on the query. (2) \ourmethod supports spatio-temporal scheduling, where a generative Scheduler is devised to evolve the dialogue-wise spatial communication topologies. (3) The Scheduler is driven by entropy and experience to perceive the system states and learn from accumulated experience. (4) The Scheduler is optimized on dozens of samples to internalize the state-aware capability and experience, thus supporting precise spatio-temporal scheduling. 

\subsection{Constructing Multi-agent Systems}
Given a query $\mathcal{Q}$, \ourmethod can adaptively build an effective MAS to handle it. Specifically, it first initializes an anchor MAS with a pre-defined communication topology $\mathcal{G}_0=(\mathcal{V},\mathcal{E}_0, \mathrm{A}_0)$. It begins by instantiating agents in $\mathcal{V}$. For each agent $v_i$, the base model, profile, and tool are assigned to it, as defined in Equation~(\ref{eq1}). To further characterize the MAS, we utilize the agents' descriptions to construct the representative features of the nodes:
\begin{gather}
    \mathcal{T}_i = \text{Text}(\text{LLM}_i + \text{Profile}_i + \text{Tool}_i),\\
    x_i^{\text{node}} = \text{NodeEncoder}(\mathcal{T}_i),
\end{gather}
where $\mathcal{T}_i$ is the textual description of agent $v_i$, and $x_i^{\text{node}}$ is the node feature generated by a sentence transformer~\citep{reimers1908sentence}, which can transform agents' descriptions into the fix-length embeddings. We denote the agent node features as $X^{\text{node}} = [x^{\text{node}}_1, x^{\text{node}}_2, \cdots, x^{\text{node}}_N]$. 

We then provide an empirically favorable communication topology $\mathrm{A}_0 \in \{0,1\}^{N\times N}$ as the anchor topology. Take an instance in math reasoning tasks given 4 agents: Manager, Analyzer, Mathematician, and Coder. The anchor topology can be organized as a Chain, i.e., ``Manager $\rightarrow$ Mathematician $\rightarrow$ Analyzer $\rightarrow$ Coder'', where the Manager assigns the task to  Mathematician and Analyzer to identify the essential mathematical problem and conduct a step-by-step analysis, finally the Coder solve numerical values through programming.

Taking the anchor communication topology $\mathcal{G}_0 = (\mathcal{V}, \mathcal{E}_0, \mathrm{A}_0)$ as a good starting point, we then utilize the generative Scheduler to construct the query-adaptive communication topology:
\begin{gather}
    \mathcal{G}_1 = \text{Scheduler}(\mathcal{Q}, t=1, \mathcal{G}_0),
\end{gather}
where $\mathcal{G}_1$ is the generated communication topology for the first iteration of dialogue. We introduce the detailed design of the Scheduler as follows.

\subsection{Generative Scheduler}
As introduced in Section~\ref{sec: formalization}, given an initial anchor topology $\mathcal{G}_0$, the Spatio-Temporal Scheduling is to autoregressively plan the future topologies $[\mathcal{G}_1, \mathcal{G}_2, \cdots, \mathcal{G}_T]$ for dialogue iterations. And it needs to simultaneously ensure \ding{182} accuracy, \ding{183} stability, and \ding{184} efficiency. 

Intuitively, adopting generative mechanisms can ensure the diversity and quality of the scheduling trajectories, and avoid overfitting during training. However, most generative mechanisms like DDPM, GAN, and VAE cannot balance these three aspects. To tackle this, since the scheduling task in \ourmethod is Spatio-Temporal Graph Generation, we draw inspiration from spatio-temporal data mining, for which lightweight flow-matching networks~\citep{liusundial,wu2026aurora} are competent. 

Specifically, we devise a compact flow-matching based Scheduler to ensure the \ding{183} stability and \ding{184} efficiency. For \ding{182} accuracy, we utilize an important property of flow-matching to serve as a stochastic interpolant, which can start from a given point, instead of a random Gaussian noise, bridging the transport path between two arbitrary distributions~\citep{wu2026aurora}.  Its core architecture consists of a GCN~\citep{zhang2019graph} and an MLP-based flow-matching network (FM-Net). The FM-Net utilizes the AdaLN~\citep{peebles2023scalable} to integrate the condition $\mathbf{H}_t$. In $t$-th dialogue, given the query $\mathcal{Q}$ and topology $\mathcal{G}_{t-1}=(\mathcal{V}, \mathcal{E}_{t-1}, \mathrm{A}_{t-1})$, the condition $\mathbf{H}_t$ is:
\begin{gather}
X^{\text{query}} = \text{Embedding}
(\mathcal{Q}),\\
X^{\text{pos}} = \text{Sinusoidal}(t), \\
\mathbf{H}_t = X^{\text{query}} + X^{\text{pos}} \in \mathbb{R}^{d}
\end{gather}
It enables the Scheduler to perceive the task information and the iteration information. Then the flow-matching process occurs in the latent space of graph embeddings, with $\mathcal{G}_{t-1}$ as the starting point:
\begin{gather}
    L_{t-1} = \text{GCN}(X^{\text{node}}, \mathrm{A}_{t-1}) + \epsilon,\\
    L_{t-(1 - p)} = p\cdot L_t + (1-p)\cdot L_{t-1},\\
    \hat{V_p} = \text{FM-Net}(L_{t-(1-p)},p,\mathbf{H}_t),
\end{gather}
where $L_{t-1}, L_{t} \in \mathbb{R}^{N\times d}$ are the graph embeddings. $L_{t-1}$ denotes the embedding of the starting point $\mathcal{G}_{t-1}$, and $L_t$ is the expected embedding of the end point $\mathcal{G}_{t}$ (we introduce this in Section~\ref{sec: scheduling}). $\epsilon \sim \mathcal{N}(\mathbf{0},\mathbf{I})$ is a Gaussian noise to increase the diversity during training. The goal of Flow-Matching (FM-Net) is: given current position $L_{t-(1-p)}$ and condition $\mathbf{H}_t$, fit the velocity field between the start and end points. Note that $p \in [0,1]$, we adopt the conditional optimal transport path to model a uniform velocity field, thus the optimization objective $\mathcal{L}_{\text{reg}}$ is:
\begin{gather}
    \mathbb{E}_{p}|| \hat{V}_p\cdot W - (L_t \cdot W - L_{t-1}\cdot W) ||^2,\label{eq: 14}
\end{gather}
where $W \in \mathbb{R}^{d\times N}$ is the projection matrix to transform graph embeddings into adjacent matrices. We also use normalization, and differentiable Bernoulli
resampling~\citep{jang2016categorical} to preserve adjacent matrices $\in\{0,1\}^{N\times N}$. In the inference phase, the sampling process is a discretized integration process--see Algorithm~\ref{alg: flow matching}.

\begin{algorithm}[H]
\captionof{algorithm}{Graph Flow Matching}
\label{alg: flow matching}
\begin{algorithmic}[1]
\State Given condition $\mathbf{H}_t$, total steps $P$,
\Statex and Topology $\mathcal{G}_{t-1}=(X^{\text{node}}, \mathrm{A}_{t-1})$.
\State Sample a Gaussian noise $\epsilon \sim \mathcal{N}(\mathbf{0}, \mathbf{I})$.
\State $\Delta p = 1/P, \hat{L} = \text{GCN}(X, \mathrm{A}_{t-1}) + \epsilon$
\State $\textbf{for}\ j\ \textbf{in}\ \{0, 1 \dots, P-1\}\ \textbf{do}$
\State $\textbf{\textcolor{white}{for}}$\ $\hat{V_j} \leftarrow \text{FM-Net}(L_{t-(1-j\Delta p)},j\Delta p,\mathbf{H}_t)$
\State $\textbf{\textcolor{white}{for}}$\ $\hat{L} \leftarrow \hat{L} + \hat{V}_j\Delta p$
\State \textbf{end for}
\State \textbf{Return:} $\hat{\mathrm{A}}_t =\text{Discretize}(\hat{L} W) \in\{0,1\}^{N\times N}$
\end{algorithmic}
\label{alg}
\end{algorithm}

\subsection{Scheduling Multi-agent Systems}
\label{sec: scheduling}

After obtaining the compact generative Scheduler, \ourmethod possesses the fundamental infrastructure for spatio-temporal scheduling, while ensuring performance, flexibility, and efficiency. The next step is to support state-aware scheduling, and enhance system's self-feedback ability. 

To precisely supervise the MAS state, we draw inspiration from uncertainty estimation~\citep{mackay1992practical}, measuring the predictive entropy PE given input $x$ by calculating the conditional entropy $H_i = H(x_i | x_{<i})$:
\begin{gather}
     H_i = \sum_{x_i}\mathrm{P}_\Theta(x_i | x_{<i})\log \mathrm{P}_\Theta(x_i | x_{<i}),\\
    \text{PE} = -\frac{1}{K}\sum_{i=1}^K H(x_i | x_{<i}),
\end{gather}
where high PE signals high uncertainty. We also consider the VarEntropy (VE), which reflects the internal self-consistency in MAS:
\begin{gather}
    \text{VE} = -\text{Std}_{i=1}^K (H(x_i|x_{<i}))
\end{gather}
With the help of PE and VE, we can classify system states into fine-grained categories--see Table~\ref{tab:entropy_varentropy_states}. 

\begin{table}[!htbp]
\centering
\resizebox{0.95\columnwidth}{!}{
\begin{tabular}{c c c}
\toprule
\textbf{PE} & \textbf{VE} & \textbf{System State Description} \\ 
\midrule
Low ($\downarrow$) & Low ($\downarrow$) & \textbf{High Confidence (Deterministic)}\\ 
\midrule
High ($\uparrow$) & High ($\uparrow$) & \textbf{Conflicting Uncertainty (Branching)} \\ 
\midrule
High ($\uparrow$) & Low ($\downarrow$) & \textbf{Uniform Ignorance (Cluelessness)} \\ \midrule
Low ($\downarrow$)&	High ($\uparrow$) & Overconfident Anomaly (Rare state) \\
\bottomrule
\end{tabular}}
\caption{Taxonomy of System States based on Predictive Entropy (PE) and VarEntropy (VE).}
\label{tab:entropy_varentropy_states}
\end{table}
Based on the summary in Table~\ref{tab:entropy_varentropy_states}, we devise a quantitative indicator, \edit{named Stability Reward ($\mathcal{R}_{sta}$)}, to evaluate the stability of MAS:
\begin{gather}
    \mathcal{R}_{\text{sta}} = -(\alpha\cdot \text{PE} + \beta\cdot \frac{1}{1 + \text{VE}}), \label{eq: 18}
\end{gather}
where higher $\mathcal{R}_{\text{sta}}$ value indicates higher stability. Note that for MAS stability measurement, $\mathcal{R}_{\text{sta}}$ is calculated over the top 10\%--20\% high-entropy tokens to prevent the stability signal from being diluted by trivial, high-confidence function words. We use $\mathcal{R}_{\text{sta}}$ as a reward model in training \ourmethod to align the preference into the Scheduler. 

On the other hand, to enhance the self-feedback ability, we devise an experience-driven method. Specifically, it is designed as a retrieve-augment system to store the historical spatio-temporal scheduling trajectories, where the indices are query embeddings: $X^{\text{query}}$. The Update strategy relies on $\mathcal{R}_{\text{sta}}$ and token consumption, considering the stability and efficiency, to store the successfully executed scheduling trajectories during training. 

After several cold-start iterations, multiple trajectories are stored, and can be used in experience training. In \ourmethod, we retrieve the most similar scheduling trajectory $\hat{\mathcal{G}}$ with current $X^{\text{query}}$, which can reflect the historical scheduling preference of the similar task. Empirically, we store the latent graph embeddings $[\hat{L}_1, \hat{L}_2, \cdots, \hat{L}_{T}]$, which are convenient to training the flow-matching based Scheduler. Recalling the Equation~(\ref{eq: 14}), we surrogate the $L_{t}$ with $\hat{L}_t$ as successful experience for reference, regularizing the Scheduler. The full details can be found in Appendix~\ref{app: rag}.

\subsection{Optimizing \ourmethod}
The ultimate objective is to achieve an effective few-shot paradigm, which trains \ourmethod on several samples, and efficiently work on numerous downstream queries. Considering the Spatio-Temporal Scheduling Communication Topologies $\mathcal{G}$ for query $\mathcal{Q}$, the utility objective is:
\begin{gather}
    \arg \underset{\Theta}{\max} \mathbb{E}_{\Theta\sim \Omega}[u(\mathcal{G}(\mathcal{Q}))],\label{eq: 16}
\end{gather}
where $\Theta$ are the parameters of Scheduler, $\Omega$ are the parameter space, and $u(\cdot)$ denote the utility. Since Equation~(\ref{eq: 16}) is inherently intractable and non-differentiable due to the external API calls for LLMs, we adopt standard approaches~\citep{GDesigner,li2025assemble,zhang2025safesieve} to apply the policy gradient for optimization:
\begin{gather}
    \frac{1}{M} \sum_{m=1}^M u(\mathcal{G}^m(\mathcal{Q}))\nabla_\Theta({P(\mathcal{G}^m)}),\label{eq: 20}
\end{gather}
where $\mathcal{G}^m$ are sampled spatio-temporal scheduling trajectories, and $\{\mathcal{G}^m(\mathcal{Q})\}_{m=1}^M$ are the outputs. $P(\mathcal{G}^m)$ is the probability of $\mathcal{G}^m$ being sampled and can be calculated through $\prod_{j=1}^T P(\mathcal{G}^m_j)$. The overall training objective of \ourmethod consists of three parts, where Equation~(\ref{eq: 18}), (\ref{eq: 20}) are reward models, and Equation~(\ref{eq: 14}) is a regularization loss function. Full details can be found in Appendix~\ref{app: loss}.

\section{Experiments}
\subsection{Experimental Settings}

\paragraph{Benchmarks} We evaluate \ourmethod on 9 benchmarks: MMLU~\citep{hendrycksmeasuring} for general reasoning, GSM8K~\citep{cobbe2021training}, MultiArith~\citep{roy2015solving}, SVAMP~\citep{patel2021nlp}, and AQuA~\citep{ling2017program} for math reasoning, HumanEval~\citep{chen2021evaluating} and DS-1000~\citep{lai2023ds} for code generation, HotpotQA~\citep{yang2018hotpotqa} for web search, and DDXPlus~\citep{fansi2022ddxplus} for medical reasoning.

\begin{table*}[!htbp]
    \centering
    \renewcommand\tabcolsep{5.3pt}
    \renewcommand\arraystretch{1.1}
    \resizebox{\linewidth}{!}{
    \begin{tabular}{lcccccccccc}
    \toprule
    \rowcb
        \textbf{Method}& \textbf{MMLU} & \textbf{GSM8K} & \textbf{MultiArith} & \textbf{SVAMP} & \textbf{AQuA} & \textbf{HumanEval} & \textbf{DS-1000} & \textbf{HotpotQA} & \textbf{DDXPlus} & \textbf{Avg.} \\ \midrule
        Vanilla & 80.47  & 87.15  & 93.40  & 87.25  & 69.27  & 73.28  & 38.40  & 82.35  & 56.40  & 74.22   \\ \hline
        \multicolumn{11}{l}{\textit{\textbf{\small Single-Agent Systems}}}
        \\
        \rowcg
        CoT  & 81.85\red{1.38}  & 87.35\red{0.20}  & 93.60\red{0.20}  & 87.90\red{0.65}  & 72.40\red{3.13}  & 74.05\red{0.77}  & 42.35\red{3.95}  & 81.04\blue{1.31}  & 64.22\red{7.82}  & 76.08   \\ 
        SC & 82.95\red{2.48}  & 87.66\red{0.51}  & 94.40\red{1.00}  & 87.40\red{0.15}  & 71.85\red{2.58}  & 75.82\red{2.54} & 37.66\blue{0.74}  & 82.24\blue{0.11}  & 66.43\red{10.03}  & 76.27   \\ \hline
        \multicolumn{11}{l}{\textit{\textbf{\small Multi-Agent Systems}}}
        \\
        \rowcg
        Chain  & 82.40\red{1.93}  & 87.23\red{0.08}  & 92.88\blue{0.52}  & 87.16\blue{0.09}  & 69.85\red{0.58}  & 73.40\red{0.12}  & 38.25\blue{0.15}  & 80.77\blue{1.58}  & 65.28\red{8.88}  & 75.25   \\ 
        Tree  & 81.55\red{1.08}  & 86.35\blue{0.80} & 92.95\blue{0.45}  & 88.25\red{1.00}  & 71.44\red{2.17}  & 75.33\red{2.05} & 43.50\red{5.10}  & 81.50\blue{0.85} & 69.25\red{12.85}  & 76.68   \\ 
        \rowcg
        Star  & 79.64\blue{0.83}  & 85.25\blue{1.90}  & 93.25\blue{0.15}  & 87.60\red{0.35}  & 68.50\blue{0.77}  & 76.28\red{3.00}  & 41.82\red{3.42}  & 83.52\red{1.17}  & 68.25\red{11.85}  & 76.01   \\ 
        Complete Graph  & 82.75\red{2.28}  & 88.05\red{0.90}  & 94.73\red{1.33}  & 88.20\red{0.95}  & 70.44\red{1.17}  & 84.25\red{10.97}  & 45.05\red{6.65}  & 83.35\red{1.00}  & 70.35\red{13.95}  & 78.57   \\ 
        \rowcg
        Random Graph & 83.20\red{2.73}  & 88.25\red{1.10}  & 94.80\red{1.40}  & 87.10\blue{0.15}  & 70.16\red{0.89}  & 84.14\red{10.86}  & 47.48\red{9.08}  & 84.28\red{1.93}  & 70.44\red{14.04}  & 78.87   \\ 
        AutoGen & 83.25\red{2.78}  & 89.40\red{2.25}  & 95.05\red{1.65}  & 88.15\red{0.90}  & 71.50\red{2.23}  & 83.50\red{10.22}  & 48.55\red{10.15}  & 83.25\red{0.90}  & 71.83\red{15.43}  & 79.39   \\ 
        \rowcg
        LangGraph & 84.10\red{3.63}  & 88.75\red{1.60}  & 95.10\red{1.70}  & 88.03\red{0.78}  & 71.27\red{2.00}  & 84.66\red{11.38}  & 47.95\red{9.55}  & 84.92\red{2.57}  & 72.50\red{16.10}  & 79.70   \\ 
        LLM-Debate & 83.28\red{2.81}  & 89.95\red{2.80}  & 95.45\red{2.05}  & 88.30\red{1.05} & 73.52\red{4.25}  & 85.24\red{11.96}  & 49.26\red{10.86}  & 84.55\red{2.20}  & 71.33\red{14.93}  & 80.10   \\ \hline
        \multicolumn{11}{l}{\textit{\textbf{\small Spatial Evolving Multi-Agent Systems}}}
        \\
        \rowcg
        GPTSwarm & 83.80\red{3.33}  & 90.20\red{3.05}  & 96.28\red{2.88}  & 87.28\red{0.03}  & 75.44\red{6.17}  & 87.14\red{13.86}  & 52.35\red{13.95}  & 85.40\red{3.05}  & 75.57\red{19.17}  & 81.50   \\ 
        G-Designer  & 84.63\red{4.16}  & 93.35\red{6.20}  & 97.60\red{4.20}  & 89.52\red{2.27}  & 76.30\red{7.03}  & 89.40\red{16.12}  & 54.82\red{16.42}  & 87.66\red{5.31}  & 76.42\red{20.02}  & 83.30   \\ 
        \rowcg
        ARG-Designer & \second{85.04}\red{4.57}  & \second{94.40}\red{7.25}  & 97.35\red{3.95}  & 88.95\red{1.70}  & 79.48\red{10.21}  & 88.25\red{14.97}  & \second{55.60}\red{17.20}  & 86.45\red{4.10}  & 76.03\red{19.63}  & 83.51   \\ 
        SafeSieve & 84.65\red{4.18}  & 93.20\red{6.05}  & \second{97.80}
        \red{4.40} & \second{90.10}\red{2.85}  & 80.40\red{11.13}  & 90.15\red{16.87}  & 53.73\red{15.33}  & 87.22\red{4.87}  & \second{77.94}\red{21.54}  & \second{83.91}   \\ \hline
        \multicolumn{11}{l}{\textit{\textbf{\small Temporal Evolving Multi-Agent Systems}}} \\   
        \rowcg
        AFlow & 83.22\red{2.75}  & 89.50\red{2.35}  & 94.28\red{0.88}  & 86.48\blue{0.77} & 78.20\red{8.93}  & 89.25\red{15.97}  & 46.82\red{8.42}  & \second{90.04}\red{7.69}  & 74.25\red{17.85}  & 81.34   \\ 
        SpecReason & 84.20\red{3.73}  & 91.40\red{4.25}  & 95.40\red{2.00}  & 87.62\red{0.37}  & 81.36\red{12.09}  & \second{91.22}\red{17.94}  & 50.08\red{11.68}  & 88.29\red{5.94}  & 76.58\red{20.18}  & 82.91   \\ 
        \rowcg
        STEER & 84.40\red{3.93}  & 93.50\red{6.35}  & 95.82\red{2.42}  & 87.40\red{0.15}  & \second{82.66}\red{13.39}  & 89.02\red{15.74}  & 51.34\red{12.94}  & 89.35\red{7.00}  & 75.20\red{18.80}  & 83.19   \\ \hline
        \multicolumn{11}{l}{\textit{\textbf{\small Spatio-Temporal Evolving Multi-Agent Systems}}} \\   
        \rowcr
        \ourmethod (Ours) & \first{89.85}\red{9.38}  & \first{97.60}\red{10.45}  & \first{99.05}\red{5.65}  & \first{94.72}\red{7.47}  & \first{86.56}\red{17.29}  & \first{94.36}\red{21.08}  & \first{58.65}\red{20.25}  & \first{92.52}\red{10.17}  & \first{82.50}\red{26.10}  & \first{88.42}  \\ \bottomrule
    \end{tabular}}
    \caption{Accuracy (\%) comparison with four types baselines, including Single-Agent Systems, Multi-Agent Systems, Spatial Evolving Multi-Agent Systems, and Temporal Evolving Multi-Agent Systems. All baselines and \ourmethod are driven by \llmname{gpt-oss-120b} (Vanilla). \first{Red}: the best, \second{Blue}: the runner-up. }
    \label{tab: main results}
\end{table*}

\paragraph{Baselines} We compare \ourmethod with static approaches like COT~\citep{wei2022chain}, Self-Consistency~\citep{wangself}, Chain, Tree, Star, Complete Graph, Random Graph, AutoGen~\citep{wu2024autogen}, LangGraph~\citep{langgraph}, LLM-Debate~\citep{du2023improving}, Spatial Evolving approaches like GPTSwarm~\citep{zhuge2024gptswarm}, G-Designer~\citep{GDesigner}, ARG-Designer~\citep{li2025assemble}, and SafeSieve~\citep{zhang2025safesieve}, Temporal Evolving approaches like AFlow~\citep{zhangaflow}, SpecReason~\citep{damani2024learning} and STEER~\citep{lee2025confidence}.  

\paragraph{Implementation Details} We run \ourmethod and all baselines with \llmname{gpt‑oss‑120b}. To synthesize dialogue history and generate the final solution, we implement a summarizer agent, maintaining $T=3$ across all baseline comparisons. For the Node Encoder and textual embedding, we adopt \llmname{all-MiniLM-L6-v2} with an embedding dimension of $D=384$. For all datasets, we use 40--80 queries to train the Scheduler. And the initial anchor topology $\mathcal{G}_0$ is customized for each dataset.

\subsection{Main Results}
Results in Table~\ref{tab: main results} demonstrate that \ourmethod achieves consistent state-of-the-art performance across nine benchmarks, with average accuracy improvements (against Vanilla) from \textit{5.65\%} to \textit{26.10\%}. Compared with Temporal or Spatial Evolving Multi-Agent Systems, \ourmethod possesses more stable performance improvement when faced with heterogeneous tasks, from simple tasks on MultiArith ($5.65 \%\uparrow$) and SVAMP ($7.47\%
\uparrow$) to hard tasks on DS-1000 ($20.25\%\uparrow$) and DDXPlus ($26.10\%\uparrow$). The larger improvements on harder tasks demonstrate the effectiveness of Spatio-Temporal Scheduling.

\subsection{Method Analysis}

\begin{figure}[!htbp]
    \centering
\includegraphics[width=1\linewidth]{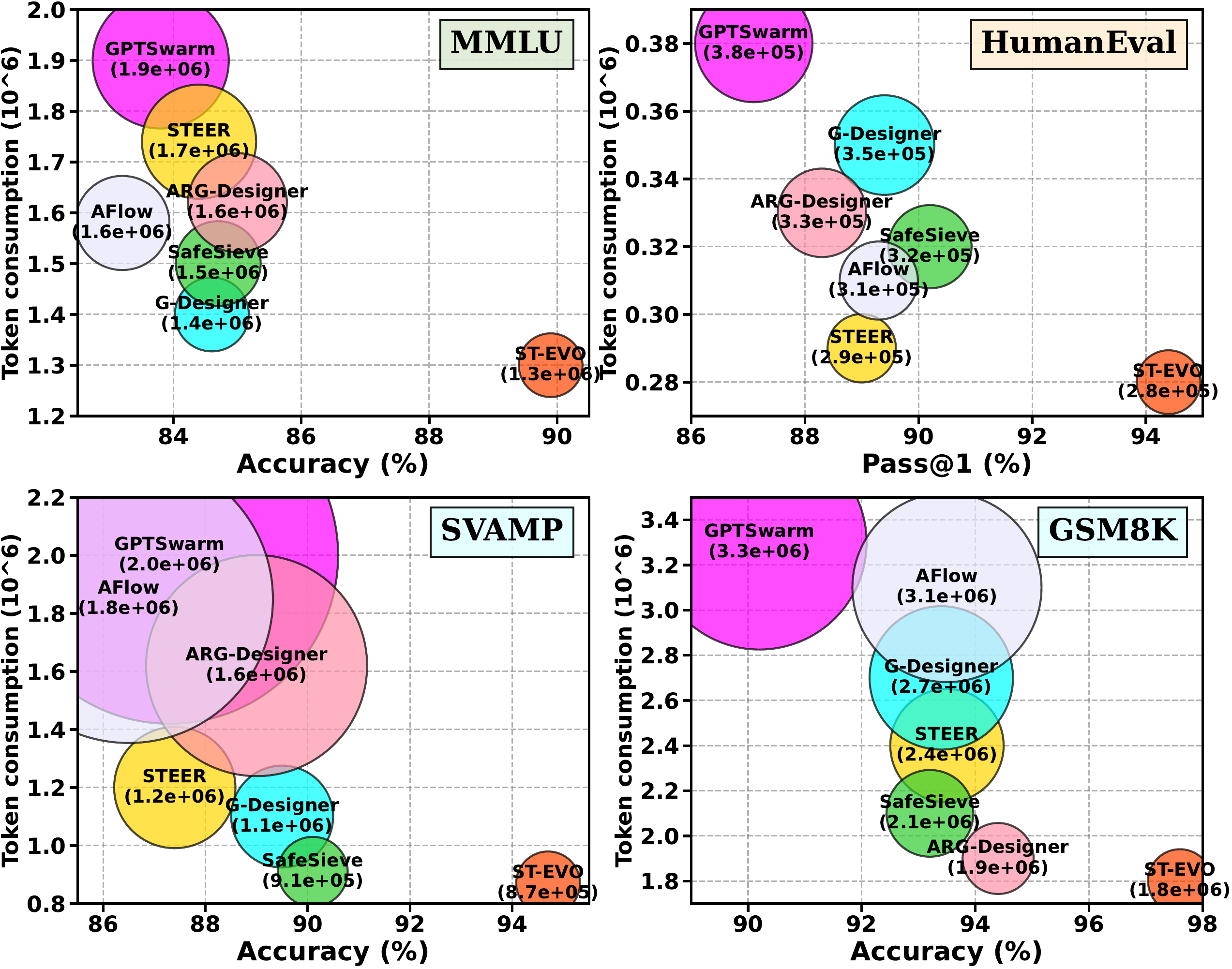}
    \caption{Visualization of performance metrics and token consumption of different MAS across MMLU, HumanEval, GSM8K, SVAMP. The diameters of the circles represent the scales of token consumption. }
\label{fig: efficiency}
\vspace{-2mm}
\end{figure}
\paragraph{Efficiency Analysis} We analyze the efficiency of \ourmethod in Figure~\ref{fig: efficiency}. When adopting the same number of agents in MAS, \ourmethod outperforms all baselines a lot with less token consumption, showing the capability to fully schedule and utilize the computing resources and specific functions in appropriate moments. Compared with previous advanced baseline ARG-Designer and SafeSieve, \ourmethod only needs nearly $50\%$ of token consumption on MMLU and HumanEval, but with higher accuracy (about $5\% \uparrow$).  We also consider the scalability in Table~\ref{tab: cross comparison} and \ref{tab: cross iteration}. Results show that the complexities of baselines like GPTSwarm and G-Designer grow linearly as the number of agents or dialogue iterations increases, while \ourmethod has a lower complexity growth rate, since spatio-temporal scheduling can effectively avoid unnecessary communication during execution.

\begin{figure}[!htbp]
    \centering
\includegraphics[width=1\linewidth]{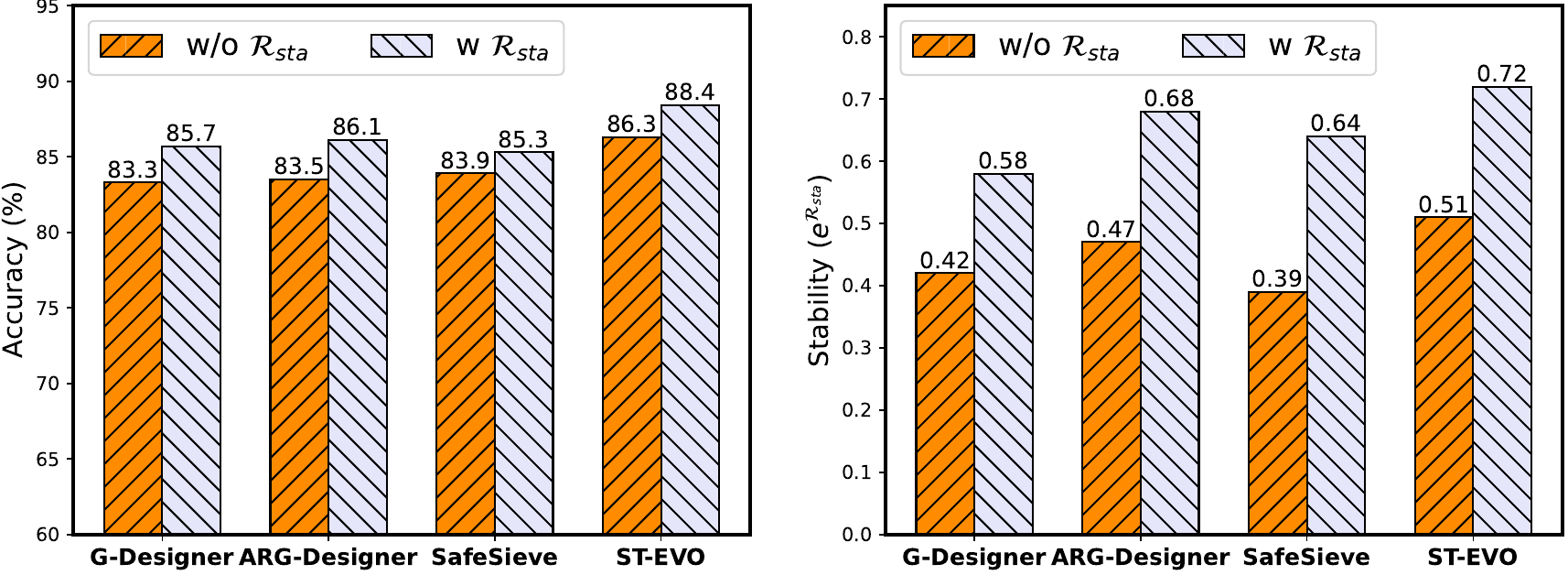}
    \caption{We utilize the accuracy (\%) and $e^{\mathcal{R}_{\text{sta}}}\in [0,1]$ to visualize the stability of \ourmethod. Higher values indicate better stabilities.}
\label{fig: stability}
\vspace{-2mm}
\end{figure}

\paragraph{Robustness Analysis} We validate \ourmethod's robustness in Figure~\ref{fig: stability}. To demonstrate whether the state-aware scheduling is useful, we first conduct experiments on all 9 benchmarks and report the average results of accuracy and stability $e^{\mathcal{R}_{\text{sta}}}$. For approaches, we select G-Designer, ARG-Designer, SafeSieve and \ourmethod since they can be trained using the \edit{Stability Reward} $\mathcal{R}_{\text{sta}}$. 

Observations show that all approaches can be improved on both the accuracy and stability using the reward $\mathcal{R}_{\text{sta}}$. This demonstrates that $\mathcal{R}_{\text{sta}}$ can effectively reflect the uncertainty of the MAS, and aligning this preference in Scheduler can lead to generalization on unseen downstream queries. Note that \ourmethod has the most improvement.

\begin{figure}[t]
    \centering
\includegraphics[width=1\linewidth]{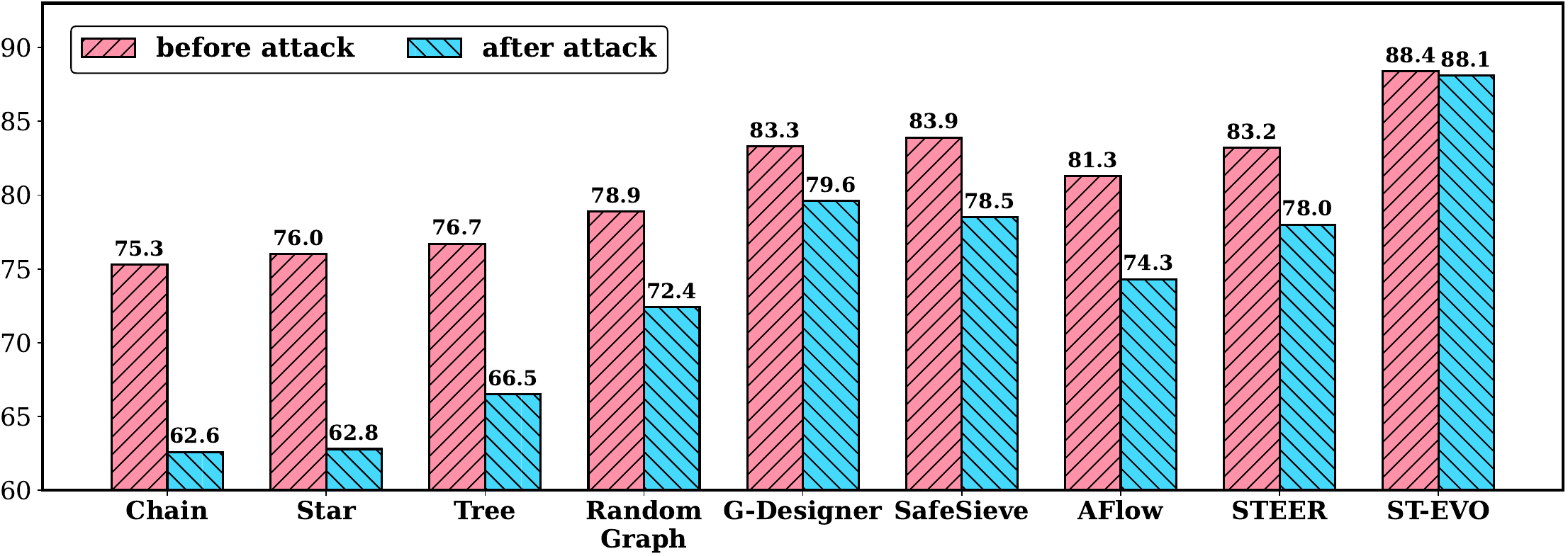}
    \caption{We compare the accuracy (\%) of multiple multi-agent systems before and after prompt attacks on all benchmarks, and report the average results.}
\label{fig: robustness}
\vspace{-2mm}
\end{figure}

We then conduct experiments to validate whether \ourmethod can fight against external disturbances. Following the experimental settings of prompt injection attack in G-Designer~\citep{GDesigner}, we compare \ourmethod with other baselines in Figure~\ref{fig: robustness}.

Results show that static multi-agent systems suffer significant accuracy degradation under partial system attacks, with drops about \textit{6.5\%--12.7\%}. Among self-evolving multi-agents, G-Designer and SafeSieve benefit from task-adaptive topology design, only suffer drops under 5\%. Thanks to such adaptive topology adjustment, \ourmethod gains exceptional robustness against adversarial attacks, and can preserve nearly consistent performance before and after the attacks.

\paragraph{Ablation Studies} We conduct ablation studies on \ourmethod--see Table~\ref{tab: ablation}: (1) w/o Scheduling, which limits the Scheduler to output a communication topology for the entire query, (2) w/o $\mathcal{L}_{\text{reg}}$, which does not rely on the experience, and (3) w/o Node Encoder, which removes the task feature $X^{\text{query}}$ in scheduling, and using adaptive representations to replace the node features.

\begin{table}[!htbp]
\centering
\resizebox{0.95\linewidth}{!}{
\begin{tabular}{l|cccc}
\toprule
Variant & MMLU & HumanEval & SVAMP & GSM8K\\ \midrule 
\ourmethod & 89.9 & 94.4 & 94.7 & 97.6 \\ \midrule
\textit{w/o} Scheduling & 84.7 & 89.6 & 88.6 & 92.5\\
\textit{w/o} $\mathcal{L}_{\text{reg}}$ & 84.9 & 88.2 & 90.3 & 93.7 \\ 
\textit{w/o} Node Encoder & 82.0 & 84.8 & 87.5 & 94.8\\
\bottomrule
\end{tabular}}
\caption{Ablation studies tested on MMLU, HumanEval, SVAMP, and GSM8K benchmarks.}
\label{tab: ablation}
\vspace{-2mm}
\end{table}
We observe that w/o Scheduling and w/o Node Encoder matters a lot, because the lack of them would break the core query-adaptive Spatio-Temporal Scheduling mechanism of \ourmethod. w/o $\mathcal{L}_{\text{reg}}$ may also make the MAS vulnerable to interference, thus degrading the performance.

\section{Conclusion}
In this work, we introduce \ourmethod, the first Spatio-Temporal Evolving MAS, which supports query-aware and dialogue-wise communication topology scheduling. Comprehensive experiments on 9 benchmarks demonstrate the state-of-the-art performance of \ourmethod, and it can ensure both efficiency and robustness on complex scenarios.

\clearpage
\section*{Limitations}
While \ourmethod demonstrates advancement on multiple tasks, it has limitations. Since it follows the few-shot learning paradigm, the Scheduler needs to be trained on a few corresponding samples to recognize the task characteristics. However, in some real-world scenarios with scarce data, this mechanism may lose its advantages. In the future, we aim at building a fully zero-shot and effective mechanism to support construction and scheduling for multi-agent systems.

\section*{Ethical considerations}
As for ethical considerations, our study only employs standard public datasets and commonly used open-source frameworks, and we find no obvious ethical issues in this work. 

\section*{Acknowledgments}
This work was partially supported by National Natural Science Foundation of China (62372179). Xingjian Wu, Xvyuan Liu, and Junkai Lu contributed equally to this work, and Bin Yang is the corresponding author. We use LLMs to conduct grammar checking and polishing on our manually-written contents.

\bibliography{reference}

@misc{langgraph,
  author = {LangChain Inc.},
  title = {LangGraph},
  year = {2024},
  howpublished = {\url{https://github.com/langchain-ai/langgraph}}
}

@misc{autogpt,
  author       = {{Significant Gravitas}},
  title        = {AutoGPT},
  year         = {2023},
  howpublished = {\url{https://github.com/Significant-Gravitas/AutoGPT}}
}

@article{wu2024autogen,
  author       = {Qingyun Wu and
                  Gagan Bansal and
                  Jieyu Zhang and
                  Yiran Wu and
                  Shaokun Zhang and
                  Erkang Zhu and
                  Beibin Li and
                  Li Jiang and
                  Xiaoyun Zhang and
                  Chi Wang},
  title        = {AutoGen: Enabling Next-Gen {LLM} Applications via Multi-Agent Conversation
                  Framework},
  journal      = {CoRR},
  volume       = {abs/2308.08155},
  year         = {2023},
  url          = {https://doi.org/10.48550/arXiv.2308.08155},
  doi          = {10.48550/ARXIV.2308.08155},
  eprinttype   = {arXiv},
  eprint       = {2308.08155},
  bibsource    = {dblp computer science bibliography, https://dblp.org}
}

@article{li2025search,
  title={Search-o1: Agentic search-enhanced large reasoning models},
  author={Li, Xiaoxi and Dong, Guanting and Jin, Jiajie and Zhang, Yuyao and Zhou, Yujia and Zhu, Yutao and Zhang, Peitian and Dou, Zhicheng},
  journal={arXiv preprint arXiv:2501.05366},
  year={2025}
}

@article{zheng2025deepresearcher,
  title={Deepresearcher: Scaling deep research via reinforcement learning in real-world environments},
  author={Zheng, Yuxiang and Fu, Dayuan and Hu, Xiangkun and Cai, Xiaojie and Ye, Lyumanshan and Lu, Pengrui and Liu, Pengfei},
  journal={arXiv preprint arXiv:2504.03160},
  year={2025}
}

@inproceedings{zheng2023codegeex,
  author       = {Qinkai Zheng and
                  Xiao Xia and
                  Xu Zou and
                  Yuxiao Dong and
                  Shan Wang and
                  Yufei Xue and
                  Lei Shen and
                  Zihan Wang and
                  Andi Wang and
                  Yang Li and
                  Teng Su and
                  Zhilin Yang and
                  Jie Tang},
  editor       = {Ambuj K. Singh and
                  Yizhou Sun and
                  Leman Akoglu and
                  Dimitrios Gunopulos and
                  Xifeng Yan and
                  Ravi Kumar and
                  Fatma Ozcan and
                  Jieping Ye},
  title        = {CodeGeeX: {A} Pre-Trained Model for Code Generation with Multilingual
                  Benchmarking on HumanEval-X},
  booktitle    = {Proceedings of the 29th {ACM} {SIGKDD} Conference on Knowledge Discovery
                  and Data Mining, {KDD} 2023, Long Beach, CA, USA, August 6-10, 2023},
  pages        = {5673--5684},
  publisher    = {{ACM}},
  year         = {2023},
  url          = {https://doi.org/10.1145/3580305.3599790},
  doi          = {10.1145/3580305.3599790},
  bibsource    = {dblp computer science bibliography, https://dblp.org}
}

@article{jin2023surrealdriver,
  title={Surrealdriver: Designing generative driver agent simulation framework in urban contexts based on large language model},
  author={Jin, Ye and Shen, Xiaoxi and Peng, Huiling and Liu, Xiaoan and Qin, Jingli and Li, Jiayang and Xie, Jintao and Gao, Peizhong and Zhou, Guyue and Gong, Jiangtao},
  journal={arXiv preprint arXiv:2309.13193},
  volume={5},
  number={7},
  pages={8},
  year={2023}
}

@inproceedings{zhuge2024gptswarm,
  title={Gptswarm: Language agents as optimizable graphs},
  author={Zhuge, Mingchen and Wang, Wenyi and Kirsch, Louis and Faccio, Francesco and Khizbullin, Dmitrii and Schmidhuber, J{\"u}rgen},
  booktitle={Forty-first International Conference on Machine Learning},
  year={2024}
}

@inproceedings{GDesigner,
  title={G-Designer: Architecting Multi-agent Communication Topologies via Graph Neural Networks},
  author={Zhang, Guibin and Yue, Yanwei and Sun, Xiangguo and Wan, Guancheng and Yu, Miao and Fang, Junfeng and Wang, Kun and Chen, Tianlong and Cheng, Dawei},
  booktitle={Forty-second International Conference on Machine Learning}
}

@misc{TimeART,
      title={TimeART: Towards Agentic Time Series Reasoning via Tool-Augmentation}, 
      author={Xingjian Wu and Junkai Lu and Zhengyu Li and Xiangfei Qiu and Jilin Hu and Chenjuan Guo and Christian S. Jensen and Bin Yang},
      year={2026},
      archivePrefix={arXiv},
      primaryClass={cs.LG}
}

@article{chen2021evaluating,
  title={Evaluating large language models trained on code},
  author={Chen, Mark},
  journal={arXiv preprint arXiv:2107.03374},
  year={2021}
}

@inproceedings{du2023improving,
  title={Improving factuality and reasoning in language models through multiagent debate},
  author={Du, Yilun and Li, Shuang and Torralba, Antonio and Tenenbaum, Joshua B and Mordatch, Igor},
  booktitle={Forty-first International Conference on Machine Learning},
  year={2023}
}

@inproceedings{wangself,
  title={Self-Consistency Improves Chain of Thought Reasoning in Language Models},
  author={Wang, Xuezhi and Wei, Jason and Schuurmans, Dale and Le, Quoc V and Chi, Ed H and Narang, Sharan and Chowdhery, Aakanksha and Zhou, Denny},
  booktitle={The Eleventh International Conference on Learning Representations},
  year={2023}
}

@article{fansi2022ddxplus,
  title={Ddxplus: A new dataset for automatic medical diagnosis},
  author={Fansi Tchango, Arsene and Goel, Rishab and Wen, Zhi and Martel, Julien and Ghosn, Joumana},
  journal={Advances in neural information processing systems},
  volume={35},
  pages={31306--31318},
  year={2022}
}

@inproceedings{yang2018hotpotqa,
  title={HotpotQA: A dataset for diverse, explainable multi-hop question answering},
  author={Yang, Zhilin and Qi, Peng and Zhang, Saizheng and Bengio, Yoshua and Cohen, William and Salakhutdinov, Ruslan and Manning, Christopher D},
  booktitle={Proceedings of the 2018 conference on empirical methods in natural language processing},
  pages={2369--2380},
  year={2018}
}

@inproceedings{lai2023ds,
  title={DS-1000: A natural and reliable benchmark for data science code generation},
  author={Lai, Yuhang and Li, Chengxi and Wang, Yiming and Zhang, Tianyi and Zhong, Ruiqi and Zettlemoyer, Luke and Yih, Wen-tau and Fried, Daniel and Wang, Sida and Yu, Tao},
  booktitle={International Conference on Machine Learning},
  pages={18319--18345},
  year={2023},
  organization={PMLR}
}

@inproceedings{ling2017program,
  title={Program Induction by Rationale Generation: Learning to Solve and Explain Algebraic Word Problems},
  author={Ling, Wang and Yogatama, Dani and Dyer, Chris and Blunsom, Phil},
  booktitle={Proceedings of the 55th Annual Meeting of the Association for Computational Linguistics (Volume 1: Long Papers)},
  pages={158--167},
  year={2017}
}

@article{patel2021nlp,
  title={Are NLP models really able to solve simple math word problems?},
  author={Patel, Arkil and Bhattamishra, Satwik and Goyal, Navin},
  journal={arXiv preprint arXiv:2103.07191},
  year={2021}
}

@inproceedings{roy2015solving,
  title={Solving general arithmetic word problems},
  author={Roy, Subhro and Roth, Dan},
  booktitle={Proceedings of the 2015 conference on empirical methods in natural language processing},
  pages={1743--1752},
  year={2015}
}

@article{cobbe2021training,
  author       = {Karl Cobbe and
                  Vineet Kosaraju and
                  Mohammad Bavarian and
                  Mark Chen and
                  Heewoo Jun and
                  Lukasz Kaiser and
                  Matthias Plappert and
                  Jerry Tworek and
                  Jacob Hilton and
                  Reiichiro Nakano and
                  Christopher Hesse and
                  John Schulman},
  title        = {Training Verifiers to Solve Math Word Problems},
  journal      = {CoRR},
  volume       = {abs/2110.14168},
  year         = {2021},
  url          = {https://arxiv.org/abs/2110.14168},
  eprinttype   = {arXiv},
  eprint       = {2110.14168},
  bibsource    = {dblp computer science bibliography, https://dblp.org}
}

@inproceedings{hendrycksmeasuring,
  title={Measuring Massive Multitask Language Understanding},
  author={Hendrycks, Dan and Burns, Collin and Basart, Steven and Zou, Andy and Mazeika, Mantas and Song, Dawn and Steinhardt, Jacob},
  booktitle={International Conference on Learning Representations}
}

@article{mackay1992practical,
  title={A practical Bayesian framework for backpropagation networks},
  author={MacKay, David JC},
  journal={Neural computation},
  volume={4},
  number={3},
  pages={448--472},
  year={1992},
  publisher={MIT Press One Rogers Street, Cambridge, MA 02142-1209, USA journals-info~…}
}

@article{jang2016categorical,
  title={Categorical reparameterization with gumbel-softmax},
  author={Jang, Eric and Gu, Shixiang and Poole, Ben},
  journal={arXiv preprint arXiv:1611.01144},
  year={2016}
}

@inproceedings{peebles2023scalable,
  title={Scalable diffusion models with transformers},
  author={Peebles, William and Xie, Saining},
  booktitle={Proceedings of the IEEE/CVF international conference on computer vision},
  pages={4195--4205},
  year={2023}
}

@article{zhang2019graph,
  title={Graph convolutional networks: a comprehensive review},
  author={Zhang, Si and Tong, Hanghang and Xu, Jiejun and Maciejewski, Ross},
  journal={Computational Social Networks},
  volume={6},
  number={1},
  pages={1--23},
  year={2019},
  publisher={Springer}
}

@inproceedings{liusundial,
  title={Sundial: A Family of Highly Capable Time Series Foundation Models},
  author={Liu, Yong and Qin, Guo and Shi, Zhiyuan and Chen, Zhi and Yang, Caiyin and Huang, Xiangdong and Wang, Jianmin and Long, Mingsheng},
  booktitle={Forty-second International Conference on Machine Learning},
  year={2025}
}

@article{ding2024large,
  title={Large language model agent in financial trading: A survey},
  author={Ding, Han and Li, Yinheng and Wang, Junhao and Chen, Hang},
  journal={arXiv preprint arXiv:2408.06361},
  year={2024}
}

@inproceedings{reimers1908sentence,
  author       = {Nils Reimers and
                  Iryna Gurevych},
  editor       = {Kentaro Inui and
                  Jing Jiang and
                  Vincent Ng and
                  Xiaojun Wan},
  title        = {Sentence-BERT: Sentence Embeddings using Siamese BERT-Networks},
  booktitle    = {Proceedings of the 2019 Conference on Empirical Methods in Natural
                  Language Processing and the 9th International Joint Conference on
                  Natural Language Processing, {EMNLP-IJCNLP} 2019, Hong Kong, China,
                  November 3-7, 2019},
  pages        = {3980--3990},
  publisher    = {Association for Computational Linguistics},
  year         = {2019},
  url          = {https://doi.org/10.18653/v1/D19-1410},
  doi          = {10.18653/V1/D19-1410},
  bibsource    = {dblp computer science bibliography, https://dblp.org}
}

@article{han2024llm,
  title={LLM multi-agent systems: Challenges and open problems},
  author={Han, Shanshan and Zhang, Qifan and Yao, Yuhang and Jin, Weizhao and Xu, Zhaozhuo},
  journal={arXiv preprint arXiv:2402.03578},
  year={2024}
}

@inproceedings{zhangaflow,
  author       = {Jiayi Zhang and
                  Jinyu Xiang and
                  Zhaoyang Yu and
                  Fengwei Teng and
                  Xionghui Chen and
                  Jiaqi Chen and
                  Mingchen Zhuge and
                  Xin Cheng and
                  Sirui Hong and
                  Jinlin Wang and
                  Bingnan Zheng and
                  Bang Liu and
                  Yuyu Luo and
                  Chenglin Wu},
  title        = {AFlow: Automating Agentic Workflow Generation},
  booktitle    = {The Thirteenth International Conference on Learning Representations,
                  {ICLR} 2025, Singapore, April 24-28, 2025},
  publisher    = {OpenReview.net},
  year         = {2025},
  url          = {https://openreview.net/forum?id=z5uVAKwmjf},
  bibsource    = {dblp computer science bibliography, https://dblp.org}
}

@article{li2025assemble,
  title={Assemble your crew: Automatic multi-agent communication topology design via autoregressive graph generation},
  author={Li, Shiyuan and Liu, Yixin and Wen, Qingsong and Zhang, Chengqi and Pan, Shirui},
  journal={arXiv preprint arXiv:2507.18224},
  year={2025}
}

@article{damani2024learning,
  title={Learning how hard to think: Input-adaptive allocation of lm computation},
  author={Damani, Mehul and Shenfeld, Idan and Peng, Andi and Bobu, Andreea and Andreas, Jacob},
  journal={arXiv preprint arXiv:2410.04707},
  year={2024}
}

@article{lee2025confidence,
  title={Confidence-Guided Stepwise Model Routing for Cost-Efficient Reasoning},
  author={Lee, Sangmook and Kim, Dohyung and Koh, Hyukhun and Yang, Nakyeong and Jung, Kyomin},
  journal={arXiv preprint arXiv:2511.06190},
  year={2025}
}

@inproceedings{wei2022chain,
  author       = {Jason Wei and
                  Xuezhi Wang and
                  Dale Schuurmans and
                  Maarten Bosma and
                  Brian Ichter and
                  Fei Xia and
                  Ed H. Chi and
                  Quoc V. Le and
                  Denny Zhou},
  editor       = {Sanmi Koyejo and
                  S. Mohamed and
                  A. Agarwal and
                  Danielle Belgrave and
                  K. Cho and
                  A. Oh},
  title        = {Chain-of-Thought Prompting Elicits Reasoning in Large Language Models},
  booktitle    = {Advances in Neural Information Processing Systems 35: Annual Conference
                  on Neural Information Processing Systems 2022, NeurIPS 2022, New Orleans,
                  LA, USA, November 28 - December 9, 2022},
  year         = {2022},
  url          = {http://papers.nips.cc/paper\_files/paper/2022/hash/9d5609613524ecf4f15af0f7b31abca4-Abstract-Conference.html},
  bibsource    = {dblp computer science bibliography, https://dblp.org}
}

@inproceedings{hong2023metagpt,
  author       = {Sirui Hong and
                  Mingchen Zhuge and
                  Jonathan Chen and
                  Xiawu Zheng and
                  Yuheng Cheng and
                  Jinlin Wang and
                  Ceyao Zhang and
                  Zili Wang and
                  Steven Ka Shing Yau and
                  Zijuan Lin and
                  Liyang Zhou and
                  Chenyu Ran and
                  Lingfeng Xiao and
                  Chenglin Wu and
                  J{\"{u}}rgen Schmidhuber},
  title        = {MetaGPT: Meta Programming for {A} Multi-Agent Collaborative Framework},
  booktitle    = {The Twelfth International Conference on Learning Representations,
                  {ICLR} 2024, Vienna, Austria, May 7-11, 2024},
  publisher    = {OpenReview.net},
  year         = {2024},
  url          = {https://openreview.net/forum?id=VtmBAGCN7o},
  bibsource    = {dblp computer science bibliography, https://dblp.org}
}

@article{yao2023tree,
  title={Tree of thoughts: Deliberate problem solving with large language models, 2023},
  author={Yao, Shunyu and Yu, Dian and Zhao, Jeffrey and Shafran, Izhak and Griffiths, Thomas L and Cao, Yuan and Narasimhan, Karthik},
  journal={URL https://arxiv. org/abs/2305.10601},
  volume={3},
  pages={1},
  year={2023}
}

@inproceedings{yao2022react,
  title={React: Synergizing reasoning and acting in language models},
  author={Yao, Shunyu and Zhao, Jeffrey and Yu, Dian and Du, Nan and Shafran, Izhak and Narasimhan, Karthik R and Cao, Yuan},
  booktitle={The eleventh international conference on learning representations},
  year={2022}
}

@article{niu2025flow,
  title={Flow: Modularized agentic workflow automation},
  author={Niu, Boye and Song, Yiliao and Lian, Kai and Shen, Yifan and Yao, Yu and Zhang, Kun and Liu, Tongliang},
  journal={arXiv preprint arXiv:2501.07834},
  year={2025}
}

@article{zhang2025safesieve,
  title={SafeSieve: From Heuristics to Experience in Progressive Pruning for LLM-based Multi-Agent Communication},
  author={Zhang, Ruijia and Zhao, Xinyan and Wang, Ruixiang and Chen, Sigen and Zhang, Guibin and Zhang, An and Wang, Kun and Wen, Qingsong},
  journal={arXiv preprint arXiv:2508.11733},
  year={2025}
}

@article{fang2025comprehensive,
  author       = {Jinyuan Fang and
                  Yanwen Peng and
                  Xi Zhang and
                  Yingxu Wang and
                  Xinhao Yi and
                  Guibin Zhang and
                  Yi Xu and
                  Bin Wu and
                  Siwei Liu and
                  Zihao Li and
                  Zhaochun Ren and
                  Nikos Aletras and
                  Xi Wang and
                  Han Zhou and
                  Zaiqiao Meng},
  title        = {A Comprehensive Survey of Self-Evolving {AI} Agents: {A} New Paradigm
                  Bridging Foundation Models and Lifelong Agentic Systems},
  journal      = {CoRR},
  volume       = {abs/2508.07407},
  year         = {2025},
  url          = {https://doi.org/10.48550/arXiv.2508.07407},
  doi          = {10.48550/ARXIV.2508.07407},
  eprinttype   = {arXiv},
  eprint       = {2508.07407},
  bibsource    = {dblp computer science bibliography, https://dblp.org}
}

@inproceedings{wang2025evoagentx,
  title={Evoagentx: An automated framework for evolving agentic workflows},
  author={Wang, Yingxu and Liu, Siwei and Fang, Jinyuan and Meng, Zaiqiao},
  booktitle={Proceedings of the 2025 Conference on Empirical Methods in Natural Language Processing: System Demonstrations},
  pages={643--655},
  year={2025}
}

@article{liu2023dynamic,
  title={Dynamic llm-agent network: An llm-agent collaboration framework with agent team optimization},
  author={Liu, Zijun and Zhang, Yanzhe and Li, Peng and Liu, Yang and Yang, Diyi},
  journal={arXiv preprint arXiv:2310.02170},
  year={2023}
}

@article{yang2024survey,
  author       = {Yiyuan Yang and
                  Ming Jin and
                  Haomin Wen and
                  Chaoli Zhang and
                  Yuxuan Liang and
                  Lintao Ma and
                  Yi Wang and
                  Chenghao Liu and
                  Bin Yang and
                  Zenglin Xu and
                  Shirui Pan and
                  Qingsong Wen},
  title        = {A Survey on Diffusion Models for Time Series and Spatio-Temporal Data},
  journal      = {{ACM} Comput. Surv.},
  volume       = {58},
  number       = {8},
  pages        = {196:1--196:39},
  year         = {2026},
  url          = {https://doi.org/10.1145/3783986},
  doi          = {10.1145/3783986},
  bibsource    = {dblp computer science bibliography, https://dblp.org}
}

@article{wang2020deep,
  title={Deep learning for spatio-temporal data mining: A survey},
  author={Wang, Senzhang and Cao, Jiannong and Philip, S Yu},
  journal={IEEE transactions on knowledge and data engineering},
  volume={34},
  number={8},
  pages={3681--3700},
  year={2020},
  publisher={IEEE}
}

@inproceedings{FlowMatch,
  author       = {Yaron Lipman and
                  Ricky T. Q. Chen and
                  Heli Ben{-}Hamu and
                  Maximilian Nickel and
                  Matthew Le},
  title        = {Flow Matching for Generative Modeling},
  booktitle    = {The Eleventh International Conference on Learning Representations,
                  {ICLR} 2023, Kigali, Rwanda, May 1-5, 2023},
  publisher    = {OpenReview.net},
  year         = {2023},
  bibsource    = {dblp computer science bibliography, https://dblp.org}
}

@inproceedings{wu2026aurora,
  title     = {Aurora: Towards Universal Generative Multimodal Time Series Forecasting},
  author    = {Wu, Xingjian and Jin, Jianxin and Qiu, Wanghui and Chen, Peng and Shu, Yang and Yang, Bin and Guo, Chenjuan},
  booktitle = {ICLR},
  year      = {2026}
}

@article{zhang2025agent,
  author       = {Kai Zhang and
                  Xiangchao Chen and
                  Bo Liu and
                  Tianci Xue and
                  Zeyi Liao and
                  Zhihan Liu and
                  Xiyao Wang and
                  Yuting Ning and
                  Zhaorun Chen and
                  Xiaohan Fu and
                  Jian Xie and
                  Yuxuan Sun and
                  Boyu Gou and
                  Qi Qi and
                  Zihang Meng and
                  Jianwei Yang and
                  Ning Zhang and
                  Xian Li and
                  Ashish Shah and
                  Dat Huynh and
                  Hengduo Li and
                  Zi Yang and
                  Sara Cao and
                  Lawrence Jang and
                  Shuyan Zhou and
                  Jiacheng Zhu and
                  Huan Sun and
                  Jason Weston and
                  Yu Su and
                  Yifan Wu},
  title        = {Agent Learning via Early Experience},
  journal      = {CoRR},
  volume       = {abs/2510.08558},
  year         = {2025},
  url          = {https://doi.org/10.48550/arXiv.2510.08558},
  doi          = {10.48550/ARXIV.2510.08558},
  eprinttype   = {arXiv},
  eprint       = {2510.08558},
  bibsource    = {dblp computer science bibliography, https://dblp.org}
}

@inproceedings{zhou2024large,
  title={Large language model as a policy teacher for training reinforcement learning agents},
  author={Zhou, Zihao and Hu, Bin and Zhao, Chenyang and Zhang, Pu and Liu, Bin},
  booktitle={Proceedings of the Thirty-Third International Joint Conference on Artificial Intelligence},
  pages={5671--5679},
  year={2024}
}

@inproceedings{qian2024chatdev,
  author       = {Chen Qian and
                  Wei Liu and
                  Hongzhang Liu and
                  Nuo Chen and
                  Yufan Dang and
                  Jiahao Li and
                  Cheng Yang and
                  Weize Chen and
                  Yusheng Su and
                  Xin Cong and
                  Juyuan Xu and
                  Dahai Li and
                  Zhiyuan Liu and
                  Maosong Sun},
  editor       = {Lun{-}Wei Ku and
                  Andre Martins and
                  Vivek Srikumar},
  title        = {ChatDev: Communicative Agents for Software Development},
  booktitle    = {Proceedings of the 62nd Annual Meeting of the Association for Computational
                  Linguistics (Volume 1: Long Papers), {ACL} 2024, Bangkok, Thailand,
                  August 11-16, 2024},
  pages        = {15174--15186},
  publisher    = {Association for Computational Linguistics},
  year         = {2024},
  url          = {https://doi.org/10.18653/v1/2024.acl-long.810},
  doi          = {10.18653/V1/2024.ACL-LONG.810},
  bibsource    = {dblp computer science bibliography, https://dblp.org}
}

@article{estornell2024multi,
  title={Multi-LLM debate: Framework, principals, and interventions},
  author={Estornell, Andrew and Liu, Yang},
  journal={Advances in Neural Information Processing Systems},
  volume={37},
  pages={28938--28964},
  year={2024}
}

@inproceedings{wangmixture,
  title={Mixture-of-Agents Enhances Large Language Model Capabilities},
  author={Wang, Junlin and Jue, WANG and Athiwaratkun, Ben and Zhang, Ce and Zou, James},
  booktitle={The Thirteenth International Conference on Learning Representations},
  year={2025}
}

@article{tang2025chemagent,
  author       = {Xiangru Tang and
                  Tianyu Hu and
                  Muyang Ye and
                  Yanjun Shao and
                  Xunjian Yin and
                  Siru Ouyang and
                  Wangchunshu Zhou and
                  Pan Lu and
                  Zhuosheng Zhang and
                  Yilun Zhao and
                  Arman Cohan and
                  Mark Gerstein},
  title        = {ChemAgent: Self-updating Library in Large Language Models Improves
                  Chemical Reasoning},
  journal      = {CoRR},
  volume       = {abs/2501.06590},
  year         = {2025},
  url          = {https://doi.org/10.48550/arXiv.2501.06590},
  doi          = {10.48550/ARXIV.2501.06590},
  eprinttype   = {arXiv},
  eprint       = {2501.06590},
  bibsource    = {dblp computer science bibliography, https://dblp.org}
}

\clearpage
\appendix

\section{Details of Methodology}
We introduce some additional technical details in this section, related to the specific implementations.
\subsection{Details of Optimization Objectives}
\label{app: loss}

Equation~(\ref{eq: 18}), (\ref{eq: 20}), and (\ref{eq: 14}) introduce three optimization objectives of \ourmethod. Among them, Equation~(\ref{eq: 18}), (\ref{eq: 20}) are optimized as reward models. Intuitively, we merge them as $\mathcal{L}_{\text{utility}}$:
\begin{gather}
    -\frac{1}{M} \sum_{m=1}^M e^{\mathcal{R}_{\text{sta}}}\cdot u(\mathcal{G}^m(\mathcal{Q}))\cdot \nabla_\Theta({P(\mathcal{G}^m)}),
\end{gather}
where $\mathcal{R}_{\text{sta}} \in \mathbb{R}^-$, so that $e^{\mathcal{R}_{\text{sta}}} \in [0,1]$ can be treated as a confidence level, which reflects the \edit{stability} of the scheduling trajectory. And the total loss function $\mathcal{L}$ is:
\begin{gather}
    \mathcal{L} = \mathcal{L}_{\text{utility}} + \gamma \cdot \mathcal{L}_{\text{reg}}
\end{gather}

\subsection{Details of Retrieve-Augment}
\label{app: rag}

To efficiently manage temporal node states under memory constraints, we introduce our retrieval mechanism. Unlike traditional graph stores, our memory $\mathcal{M}$ with capacity $L$ stores latent tensor representations. Each item $m_i \in \mathcal{M}$ is a tuple $(\mathbf{k}_i, \mathcal{H}_i, c_i, u_i, \alpha_i)$, where $\mathbf{k}_i$ is the query embedding key, $\mathcal{H}_i$ is the sequence of node feature tensors (values), $c_i$ is the computational cost, $u_i$ is the \edit{Stability Reward} (i.e., $\mathcal{R}_{\text{sta}}$), and $\alpha_i$ tracks access frequency.

\textbf{Retrieval and Update.} 
Given a query $\mathbf{q}$, we retrieve the top-$k$ items maximizing the dot-product similarity $\mathbf{k}_i^\top \mathbf{q}$. To maintain the most valuable information within the fixed capacity $L$, we employ a value-based eviction policy. When the buffer is full, a new candidate $m_{new}$ replaces the item with the minimum retention score $S(m)$ if $S(m_{new}) > \min_{j} S(m_j)$. The score $S(m)$ balances reliability, efficiency, and popularity:

\begin{equation}
    S(m_i) = \frac{1 + \log(\alpha_i + 1)}{c_i \cdot |u_i| + \epsilon}
\end{equation}

where $\epsilon$ is a smoothing term. This policy prioritizes items with \edit{high stability} ($|u_i| \to 0$) and low computational cost ($c_i$), while logarithmically scaling with access frequency to preserve useful knowledge without saturation.

\subsection{Entropy Estimation without Token Log-Probabilities}
\label{app:entropy_proxy}

The original stability reward uses token-level log-probabilities to compute Predictive Entropy (PE) and VarEntropy (VE). When an inference backend does not expose token probabilities, we use two alternative estimators. The first is a sample-based estimator that draws $K$ responses for the same prompt. Let $p(a)$ denote the empirical frequency of answer $a$ among these responses. We estimate answer-level uncertainty as
\begin{equation}
    \widehat{\mathrm{PE}}=-\sum_a p(a)\log p(a).
\end{equation}
We estimate response-level variation from the mean pairwise cosine distance between response embeddings. Given embeddings $\{\mathbf{e}_i\}_{i=1}^{K}$, this proxy is
\begin{equation}
    \widehat{\mathrm{VE}}=1-\frac{2}{K(K-1)}
    \sum_{i<j}\cos(\mathbf{e}_i,\mathbf{e}_j).
\end{equation}
The second alternative is a lightweight confidence probe: a small MLP predicts PE and VE directly from output-text features and therefore requires neither token probabilities nor repeated API calls.

\begin{table}[!htbp]
\centering
\small
\resizebox{\columnwidth}{!}{
\begin{tabular}{lccc}
\toprule
\textbf{Entropy Estimator} & \textbf{Avg. Acc.} & \textbf{PE Corr.} & \textbf{Extra Calls} \\
\midrule
Logprob-based & \textbf{94.15} & 1.00 & 0 \\
Sample-based ($K=5$) & 92.95 & 0.85 & $5\times$ \\
Sample-based ($K=3$) & 92.23 & -- & $3\times$ \\
Confidence probe & 92.40 & 0.78 & 0 \\
w/o PE/VE & 91.80 & -- & 0 \\
\bottomrule
\end{tabular}}
\caption{Comparison of entropy estimators. Average accuracy is computed over MMLU, HumanEval, SVAMP, and GSM8K. PE Corr. is the Pearson correlation with logprob-based PE, and extra calls are measured per agent and iteration.}
\label{tab:entropy_proxy}
\end{table}

Table~\ref{tab:entropy_proxy} shows that sample-based estimation with $K=5$ is only $1.20$ points below the logprob-based implementation and remains $1.15$ points above removing entropy altogether. The confidence probe also preserves most of the benefit without additional model calls. Thus, token probabilities provide the most accurate and efficient signal when available, but they are not a strict deployment requirement for \ourmethod.

\section{Details of Experiments}

\subsection{Dataset Statistics}
\begin{table}[!htbp]
\centering
\resizebox{\linewidth}{!}{
\begin{tabular}{llllll}
\toprule
\textbf{Category} & \textbf{Dataset} & \textbf{Answer Type} & \textbf{Metric} & \textbf{\#Test} & \textbf{License} \\
\midrule
\multirow{2}{*}{General reasoning} & MMLU & Multi-choice & Acc. & 153 & MIT License \\
 & HotpotQA & Span & F1 & 7,405 & CC BY-SA 4.0 \\
\midrule
\multirow{4}{*}{Math reasoning} & GSM8K & Number & Acc. & 1,319 & MIT License \\
 & MultiArith & Number & Acc. & 600 & Unspecified \\
 & SVAMP & Number & Acc. & 1,000 & MIT License \\
 & AQuA & Multi-choice & Acc. & 254 & Apache-2.0 \\
\midrule
\multirow{2}{*}{Code generation} & HumanEval & Code & Pass@1 & 164 & MIT License \\
 & DS-1000 & Code & Acc. & 1,000 & CC BY-SA 4.0 \\
\midrule
Medical reasoning & DDXPlus & Multi-choice & Acc. & 282,906 & CC BY-SA 4.0 \\
\bottomrule
\end{tabular}}
\caption{Dataset descriptions and statistics.}
\label{tab: datasets}
\end{table}

To comprehensively evaluate the model's capabilities, we selected a diverse set of benchmarks across four primary categories: general reasoning, mathematical reasoning, code generation, and medical reasoning. Table~\ref{tab: datasets} provides a detailed summary of these datasets, including their respective evaluation metrics, test set sizes, and licenses.

\subsection{Evaluation on Challenging Agentic Benchmarks}
\label{app:hard_benchmarks}

The nine standard benchmarks provide controlled coverage of reasoning, code generation, retrieval, and medical tasks. To test whether dynamic coordination remains useful in more realistic long-horizon settings, we additionally evaluate on SWE-Bench Verified, Terminal-Bench, MLE-Bench, and WebArena. These benchmarks require codebase-level debugging, terminal interaction, end-to-end machine-learning engineering, and interactive web navigation, respectively. All methods use \texttt{gpt-oss-120b} under the same communication-based multi-agent protocol; consequently, the absolute numbers are not directly comparable with single-agent tool-use pipelines that use different scaffolds.

\begin{table*}[!htbp]
\centering
\small
\resizebox{0.8\textwidth}{!}{
\begin{tabular}{lcccc}
\toprule
\textbf{Method} & \textbf{SWE-Bench} & \textbf{Terminal-Bench} & \textbf{MLE-Bench} & \textbf{WebArena} \\
\midrule
Vanilla (single) & 37.4 & 32.6 & 9.2 & 15.7 \\
Chain (4 agents) & 39.1 & 34.8 & 10.6 & 17.9 \\
Complete Graph & 40.8 & 36.3 & 12.1 & 20.3 \\
G-Designer & 44.2 & 39.7 & 15.8 & 24.6 \\
SafeSieve & 45.6 & 40.9 & 16.4 & 25.8 \\
STEER & \underline{46.3} & \underline{41.5} & \underline{17.2} & \underline{26.7} \\
\ourmethod & \textbf{51.8} & \textbf{47.6} & \textbf{22.7} & \textbf{33.1} \\
\bottomrule
\end{tabular}}
\caption{Results (\%) on four challenging agentic benchmarks. All methods use \texttt{gpt-oss-120b} and the same communication-based multi-agent protocol.}
\label{tab:hard_agentic_benchmarks}
\end{table*}

As shown in Table~\ref{tab:hard_agentic_benchmarks}, \ourmethod improves over the strongest baseline by $5.5$, $6.1$, $5.5$, and $6.4$ points on the four benchmarks, respectively. These gains are at least as large as those on the standard benchmark suite, indicating that dialogue-wise topology adaptation becomes more valuable as tasks require longer planning horizons and more heterogeneous forms of collaboration.

\subsection{Generalization across Backbone Models}
\label{app:backbone_generalization}

We evaluate whether the learned scheduler depends on a particular backbone by replacing \texttt{gpt-oss-120b} with five additional self-deployable models. The models cover dense, mixture-of-experts (MoE), and code-specialized architectures, with active parameter counts ranging from 3B to 120B. Table~\ref{tab:backbone_generalization} reports averages over the original nine benchmarks.

\begin{table*}[!htbp]
\centering
\small
\resizebox{0.9\textwidth}{!}{
\begin{tabular}{llcccc}
\toprule
\textbf{Backbone} & \textbf{Architecture} & \textbf{Vanilla} & \textbf{Best Baseline} & \textbf{\ourmethod} & \textbf{Gain} \\
\midrule
\texttt{gpt-oss-120b} & 120B Dense & 74.22 & 83.91 & 88.42 & $+4.51$ \\
\texttt{Qwen3.5-122B-A10B} & 122B MoE (10B active) & 71.99 & 81.75 & 86.48 & $+4.73$ \\
\texttt{Qwen3.6-27B} & 27B Dense & 70.05 & 79.88 & 84.83 & $+4.95$ \\
\texttt{Qwen3.6-35B-A3B} & 35B MoE (3B active) & 64.16 & 75.18 & 81.14 & $+5.96$ \\
\texttt{Qwen3-30B-A3B} & 30B MoE (3B active) & 62.71 & 73.74 & 79.59 & $+5.85$ \\
\texttt{Qwen3-Coder-30B-A3B} & 30B MoE (3B active) & 61.67 & 72.50 & 78.29 & $+5.79$ \\
\bottomrule
\end{tabular}}
\caption{Average accuracy (\%) across the nine benchmarks with different backbone models. Gain is measured against the strongest baseline for each backbone.}
\label{tab:backbone_generalization}
\end{table*}

\ourmethod consistently improves over the strongest baseline across all model sizes and architectures. The gains are larger for models with only 3B active parameters ($5.79$--$5.96$ points) than for the 120B dense model ($4.51$ points), suggesting that adaptive collaboration can compensate for weaker individual agents. Across the 3B-active models, PE/VE remains correlated with output quality (Pearson $r>0.65$), supporting entropy-aware scheduling at smaller scales.

\subsection{Scalability}

\begin{table}[!htbp]
\centering
\resizebox{0.9\linewidth}{!}{
\begin{tabular}{c|ccc}
\toprule
\#Agents & 5 & 10 & 20 \\ \midrule
\textbf{Random Graph} & & & \\
Accuracy (\%) & 82.7 & 83.3 & 83.6 \\
Time (min) & 14.2 & 31.4 & 59.7 \\
\#Tokens & 431,064 & 1,317,148 & 4,245,925 \\
\midrule
\textbf{GPTSwarm} & & & \\
Accuracy (\%) & 83.6 & 84.2 & 84.8 \\
Time (min) & 67.8 & 201.8 & 422.5 \\
\#Tokens & 1,672,201 & 5,103,328 & 14,054,825  \\ \midrule
\textbf{G-Designer} & & & \\
Accuracy (\%) & 84.1 & 85.2 & 87.3 \\
Time (min) & 22.4 & 41.8 & 75.23 \\
\#Tokens & 1,422,309 & 2,887,532 & 6,130,221 \\ \midrule
\textbf{AFlow} & & & \\
Accuracy (\%) & 83.0 & 84.5 & 85.1 \\
Time (min) & 36.7 & 77.2 & 135.4 \\
\#Tokens & 1,851,925 & 3,485,925 & 7,292,296 \\ \midrule
\textbf{STEER} & & & \\
Accuracy (\%) & 84.0 & 86.2 & 87.8 \\
Time (min) & 48.2 & 105.6 & 233.2 \\
\#Tokens & 2,025,125 & 4,256,252 & 9,026,290 \\ \midrule
\textbf{\ourmethod} & & & \\
Accuracy (\%) & 89.2 & 91.6 & 93.3 \\
Time (min) & 15.2 & 28.4 & 51.6 \\
\#Tokens & 1,162,296 & 1,825,926 & 3,326,037 \\
\bottomrule

\end{tabular}}
\caption{Comparison of accuracy, time, and token consumption across different agent configurations. We use the MMLU benchmark and \llmname{gpt-oss-120b} as the base LLM.}
\label{tab: cross comparison}
\end{table}

\begin{table}[!htbp]
\centering
\resizebox{0.9\linewidth}{!}{
\begin{tabular}{c|ccc}
\toprule
\#Iterations & 3 & 6 & 12 \\ \midrule
\textbf{Random Graph} & & & \\
Accuracy (\%) & 82.7 &83.1  &83.3  \\
Time (min) & 14.2 & 27.4 & 52.5 \\
\#Tokens & 431,064 & 816,266  &1,425,295  \\
\midrule
\textbf{GPTSwarm} & & & \\
Accuracy (\%) & 83.6 &84.0  &84.2  \\
Time (min) & 67.8 & 112.5 & 199.4  \\
\#Tokens & 1,672,201 & 3,025,162  & 5,925,102   \\ \midrule
\textbf{G-Designer} & & & \\
Accuracy (\%) & 84.1 &83.8  & 84.3 \\
Time (min) & 22.4 & 42.5 & 79.2  \\
\#Tokens & 1,422,309 & 3,015,285 & 5,723,293  \\ \midrule
\textbf{AFlow} & & & \\
Accuracy (\%) & 83.0 & 83.4 & 83.6 \\
Time (min) & 36.7 & 67.2 & 122.4 \\
\#Tokens & 1,851,925 &3,425,291  &6,902,946  \\ \midrule
\textbf{STEER} & & & \\
Accuracy (\%) & 84.0 & 84.3  & 84.5 \\
Time (min) & 48.2 & 87.2 & 144.5 \\
\#Tokens & 2,025,125 & 3,821,229 & 7,402,409 \\ \midrule
\textbf{\ourmethod} & & & \\
Accuracy (\%) & 89.2 & 89.4 & 89.6\\
Time (min) & 15.2 & 26.4 & 44.2 \\
\#Tokens & 1,162,296 & 1,823,394 & 2,901,034 \\
\bottomrule

\end{tabular}}
\caption{Comparison of accuracy, time, and token consumption across different dialogue iterations. We use the MMLU benchmark and \llmname{gpt-oss-120b} as the base LLM.}
\label{tab: cross iteration}
\end{table}

\section{Ablation Study on the Scheduler}
\label{app:ablation_scheduler}

\edit{To further examine the contribution of the scheduler in \ourmethod, we evaluate different scheduling strategies and generative mechanisms, conduct combined module ablations, analyze the sensitivity to core hyperparameters, and quantify topology changes between adjacent iterations.}

\subsection{Ablation on scheduling strategies.}
\edit{We first design several simple scheduler variants for comparison. 
First, \textit{w/ LLM Scheduler} replaces our learned scheduler with a language model, where \texttt{gpt-oss-120b} directly reasons over the current dialogue state and determines the communication topology at each reasoning step. 
Second, \textit{w/o Scheduling} only uses our generated scheduler to produce the initial communication topology, and then keeps this topology fixed throughout the entire multi-turn reasoning process, which resembles the static-topology setting in G-Designer. 
Third, \textit{w/ Random Scheduler} randomly samples a communication topology at each reasoning step, without considering the evolving intermediate reasoning states.}

\begin{table}[!htbp]
\centering
\small
\setlength{\tabcolsep}{5pt}
\renewcommand{\arraystretch}{1.08}
\resizebox{\columnwidth}{!}{
\begin{tabular}{lccccc}
\toprule
Variant & MMLU & HumanEval & SVAMP & GSM8K & Avg. \\
\midrule
\textbf{\ourmethod} 
& \textbf{89.9} & \textbf{94.4} & \textbf{94.7} & \textbf{97.6} & \textbf{94.15} \\
w/ LLM Scheduler    
& 86.8 & 91.2 & 91.5 & 94.9 & 91.10 \\
w/o Scheduling      
& 84.7 & 89.6 & 88.6 & 92.5 & 88.85 \\
w/ Random Scheduler 
& 83.8 & 88.7 & 87.9 & 91.6 & 88.00 \\
\bottomrule
\end{tabular}
}
\caption{
Ablation study on different scheduling strategies. 
We report the accuracy (\%) on MMLU, HumanEval, SVAMP, and GSM8K, as well as the average performance.
}
\label{tab:ablation_scheduler_strategy}
\end{table}

\edit{As shown in Table~\ref{tab:ablation_scheduler_strategy}, \ourmethod consistently outperforms all scheduler variants across the four benchmarks. 
Replacing the learned scheduler with an LLM-based scheduler leads to an average drop of $3.05\%$, suggesting that directly prompting a language model to design the topology is insufficient for stable and fine-grained multi-step coordination. 
The \textit{w/o Scheduling} variant further degrades the average performance by $5.30\%$, indicating that using a fixed topology cannot effectively adapt to the evolving reasoning states during multi-agent collaboration. 
The random scheduler performs the worst, with an average drop of $6.15\%$, which confirms that dynamically changing the topology alone is not enough; the topology must be optimized according to the current reasoning context. 
These results demonstrate that the proposed scheduler is essential for enabling effective step-wise communication and coordination among agents.}

\subsection{Ablation on generative mechanisms}
\edit{We further study why \ourmethod adopts Flow Matching as the generative mechanism for topology scheduling. 
The key motivation is that Flow Matching can naturally serve as an interpolator: it starts from an arbitrary source distribution and learns a velocity field that transports it toward the target distribution. 
This property allows us to directly take the communication topology from the previous reasoning step as the starting point, and then learn the transition toward the target topology of the next step. 
In this way, the starting topology and the conditional information are explicitly decoupled, making the scheduler better aligned with the dynamic nature of multi-step topology evolution.}

\edit{By contrast, VAE- and Diffusion-based generative mechanisms need to inject the previous communication topology into the conditional network to achieve a similar effect. 
However, the conditional network also needs to encode task features and step-dependent features. 
Coupling the previous topology with these heterogeneous conditions may make the conditioning signal entangled and hinder the extraction of key scheduling information. 
To verify this point, we construct two additional variants. 
For \textit{w/ Diffusion Scheduler}, we adopt DDIM as the sampling mechanism, keep the same MLP-based scheduler architecture as \ourmethod, and use AdaLN for condition injection. 
The training objective is changed to a denoising objective with the negative ELBO loss. 
For \textit{w/ VAE Scheduler}, we adopt a conditional VAE decoder with the same condition-injection strategy, and train it with the negative ELBO loss.}

\begin{table}[!htbp]
\centering
\small
\setlength{\tabcolsep}{5pt}
\renewcommand{\arraystretch}{1.08}
\resizebox{\columnwidth}{!}{
\begin{tabular}{lccccc}
\toprule
Variant & MMLU & HumanEval & SVAMP & GSM8K & Avg. \\
\midrule
\textbf{\ourmethod} 
& \textbf{89.9} & \textbf{94.4} & \textbf{94.7} & \textbf{97.6} & \textbf{94.15} \\
w/ Diffusion Scheduler 
& 87.6 & 92.0 & 92.4 & 95.5 & 91.88 \\
w/ VAE Scheduler       
& 87.1 & 91.6 & 91.8 & 95.0 & 91.38 \\
\bottomrule
\end{tabular}
}
\caption{
Ablation study on different generative mechanisms for topology scheduling.
We report the accuracy (\%) on MMLU, HumanEval, SVAMP, and GSM8K, as well as the average performance.
}
\label{tab:ablation_scheduler_generation}
\end{table}

\edit{Table~\ref{tab:ablation_scheduler_generation} shows that the Flow-Matching-based scheduler achieves the best performance on all benchmarks. 
Compared with \ourmethod, the Diffusion-based scheduler suffers an average drop of $2.27\%$, while the VAE-based scheduler leads to a larger average drop of $2.77\%$. 
These results support our design choice: modeling topology evolution as a flow from the previous communication topology to the next target topology provides a more direct and effective formulation than coupling the previous topology into the conditional network of VAE- or Diffusion-based generators. 
Therefore, Flow Matching is better suited for learning communication transitions in \ourmethod.}

\subsection{Combined Ablation of Feedback Modules}
\label{app:combined_ablation}

The stability reward provides real-time state feedback, whereas retrieval provides historical scheduling experience. We jointly remove these components to determine whether one can substitute for the other.

\begin{table}[!htbp]
\centering
\small
\resizebox{\columnwidth}{!}{
\begin{tabular}{lccccc}
\toprule
\textbf{Variant} & \textbf{MMLU} & \textbf{HumanEval} & \textbf{SVAMP} & \textbf{GSM8K} & \textbf{Avg.} \\
\midrule
\ourmethod (full) & \textbf{89.90} & \textbf{94.40} & \textbf{94.70} & \textbf{97.60} & \textbf{94.15} \\
w/o Entropy Reward & 87.23 & 91.76 & 92.41 & 95.82 & 91.81 \\
w/o Retrieval & 84.86 & 88.24 & 90.31 & 93.72 & 89.28 \\
w/o Both & 83.17 & 86.43 & 87.94 & 91.76 & 87.33 \\
\bottomrule
\end{tabular}}
\caption{Combined ablation of the entropy-based stability reward and retrieve-augment module. All values are accuracy (\%).}
\label{tab:combined_feedback_ablation}
\end{table}

Removing the entropy reward and retrieval separately decreases the average by $2.34$ and $4.87$ points, respectively. Removing both produces a $6.82$-point drop, close to the $7.21$-point drop predicted by adding the two individual effects. This result indicates that the modules provide largely complementary information: entropy characterizes the current system state, while retrieval supplies relevant historical trajectories.

\subsection{Sensitivity to Core Hyperparameters}
\label{app:hyperparameter_ablation}

We vary the stability-reward weights $(\alpha,\beta)$, trajectory-cache capacity $L$, and number of Flow-Matching integration steps $P$. Unless varied, the other hyperparameters use their default values.

\begin{table}[!htbp]
\centering
\small
\resizebox{\columnwidth}{!}{
\begin{tabular}{cccccc}
\toprule
$\alpha$ & $\beta$ & \textbf{MMLU} & \textbf{HumanEval} & \textbf{SVAMP} & \textbf{GSM8K / Avg.} \\
\midrule
0.3 & 0.7 & 88.17 & 92.68 & 93.14 & 96.33 / 92.58 \\
0.5 & 0.5 & \textbf{89.85} & \textbf{94.36} & \textbf{94.72} & \textbf{97.60 / 94.13} \\
0.7 & 0.3 & 89.06 & 93.51 & 93.85 & 96.92 / 93.34 \\
1.0 & 0.0 & 87.43 & 91.74 & 91.96 & 95.48 / 91.65 \\
0.0 & 1.0 & 86.72 & 91.15 & 91.37 & 94.93 / 91.04 \\
\bottomrule
\end{tabular}}
\caption{Sensitivity to the stability-reward weights $\alpha$ and $\beta$. Results are accuracy (\%).}
\label{tab:stability_weight_ablation}
\end{table}

Balanced weights achieve the best average performance. Setting either coefficient to zero causes a drop of at least $2.48$ points, confirming that PE and VE capture complementary aspects of system uncertainty.

\begin{table}[!htbp]
\centering
\small
\resizebox{\columnwidth}{!}{
\begin{tabular}{cccccc}
\toprule
\textbf{Capacity $L$} & \textbf{MMLU} & \textbf{HumanEval} & \textbf{SVAMP} & \textbf{GSM8K} & \textbf{Avg.} \\
\midrule
0 & 84.86 & 88.24 & 90.31 & 93.72 & 89.28 \\
25 & 87.74 & 91.93 & 92.76 & 95.81 & 92.06 \\
50 & 88.93 & 93.47 & 93.85 & 96.74 & 93.25 \\
100 & \textbf{89.90} & \textbf{94.40} & \textbf{94.70} & \textbf{97.60} & \textbf{94.15} \\
200 & 89.82 & 94.31 & 94.63 & 97.54 & 94.08 \\
500 & 89.79 & 94.28 & 94.58 & 97.51 & 94.04 \\
\bottomrule
\end{tabular}}
\caption{Sensitivity to trajectory-cache capacity. Results are accuracy (\%).}
\label{tab:cache_capacity_ablation}
\end{table}

Performance improves steadily up to $L=100$ and then saturates. A moderate cache therefore captures sufficient reusable experience, while larger caches provide no material gain and may introduce less relevant trajectories.

\begin{table}[!htbp]
\centering
\small
\resizebox{\columnwidth}{!}{
\begin{tabular}{ccccccc}
\toprule
$P$ & \textbf{MMLU} & \textbf{HumanEval} & \textbf{SVAMP} & \textbf{GSM8K} & \textbf{Avg.} & \textbf{Time (ms)} \\
\midrule
10 & 86.53 & 90.74 & 91.18 & 94.46 & 90.73 & 12 \\
25 & 88.19 & 92.83 & 93.41 & 96.23 & 92.67 & 24 \\
50 & 89.24 & 93.79 & 94.05 & 96.98 & 93.52 & 38 \\
100 & 89.85 & 94.36 & 94.72 & 97.60 & 94.13 & 65 \\
200 & 89.91 & 94.38 & 94.74 & 97.62 & 94.16 & 125 \\
500 & \textbf{89.94} & \textbf{94.42} & \textbf{94.79} & \textbf{97.65} & \textbf{94.20} & 305 \\
\bottomrule
\end{tabular}}
\caption{Accuracy--latency trade-off for the number of Flow-Matching integration steps $P$.}
\label{tab:flow_sampling_steps}
\end{table}

Accuracy largely saturates at $P=100$: increasing $P$ from 100 to 500 improves the average by only $0.07$ points while increasing topology-generation time from 65 to 305 ms. We therefore use $P=100$ as the default accuracy--efficiency trade-off.

\subsection{Topological Oscillation and Smoothing}
\label{app:topology_oscillation}

We measure the similarity of adjacent communication topologies using the Jaccard similarity between their edge sets,
\begin{equation}
    J(\mathcal{E}_{t-1},\mathcal{E}_t)=
    \frac{|\mathcal{E}_{t-1}\cap\mathcal{E}_t|}
    {|\mathcal{E}_{t-1}\cup\mathcal{E}_t|}.
\end{equation}
A higher value denotes a smoother transition. To explicitly discourage abrupt changes, we optionally add an adjacency regularizer
\begin{equation}
    \mathcal{L}_{\mathrm{smooth}}=
    \lambda\|\mathrm{A}_t-\mathrm{A}_{t-1}\|_1.
\end{equation}

\begin{table}[!htbp]
\centering
\small
\resizebox{\columnwidth}{!}{
\begin{tabular}{lccc}
\toprule
\textbf{Method} & \textbf{Jaccard} & \textbf{Std.} & \textbf{Oscillation Rate} \\
\midrule
Random Scheduler & 0.31 & 0.22 & 78.4\% \\
G-Designer (fixed) & 1.00 & 0.00 & 0.0\% \\
\ourmethod w/o smoothing & 0.72 & 0.15 & 22.3\% \\
\ourmethod w/ smoothing & 0.81 & 0.09 & 12.8\% \\
\bottomrule
\end{tabular}}
\caption{Similarity and oscillation of communication topologies between adjacent dialogue iterations.}
\label{tab:topology_oscillation}
\end{table}

\begin{table}[!htbp]
\centering
\small
\resizebox{\columnwidth}{!}{
\begin{tabular}{lccccc}
\toprule
\textbf{Variant} & \textbf{MMLU} & \textbf{HumanEval} & \textbf{SVAMP} & \textbf{GSM8K} & \textbf{Jaccard} \\
\midrule
w/o smoothing & \textbf{89.85} & \textbf{94.36} & \textbf{94.72} & \textbf{97.60} & 0.72 \\
$\lambda=0.01$ & 89.78 & 94.29 & 94.64 & 97.53 & 0.78 \\
$\lambda=0.05$ & 89.71 & 94.18 & 94.51 & 97.42 & 0.81 \\
$\lambda=0.10$ & 89.32 & 93.81 & 94.03 & 97.06 & 0.88 \\
$\lambda=0.50$ & 87.46 & 91.53 & 92.18 & 95.27 & \textbf{0.95} \\
\bottomrule
\end{tabular}}
\caption{Effect of topology-smoothing strength on accuracy (\%) and adjacent-topology similarity.}
\label{tab:topology_smoothing}
\end{table}

Even without an explicit penalty, Flow Matching yields relatively gradual transitions ($J=0.72$). Small smoothing weights further reduce oscillation with less than $0.2$ points of accuracy loss, whereas excessive smoothing prevents necessary topology adaptation. We use $\lambda=0.01$ when smoother deployment behavior is preferred; otherwise, no explicit smoothing is required.

\subsection{Generalization Study via Cross-Dataset Transfer}
\label{app:generalization_transfer}

\edit{We further conduct a transfer study to examine the generalization capability of \ourmethod. 
In this experiment, $A \rightarrow B$ denotes that the scheduler is few-shot trained on the source dataset $A$ and then directly evaluated on the target dataset $B$. 
We hypothesize that both the scheduling trajectory memory and the learned scheduler parameters are transferable across tasks with similar reasoning structures. 
Therefore, we first consider two in-domain transfer settings: GSM8K $\rightarrow$ SVAMP for mathematical reasoning, and HumanEval $\rightarrow$ DS-1000 for code reasoning. 
We also evaluate two cross-domain transfer settings, including GSM8K $\rightarrow$ HumanEval and HumanEval $\rightarrow$ GSM8K, to study whether the learned scheduling ability can generalize across substantially different task domains.}

\begin{table}[t]
\centering
\small
\setlength{\tabcolsep}{4pt}
\renewcommand{\arraystretch}{1.08}
\resizebox{0.95\columnwidth}{!}{
\begin{tabular}{llccc}
\toprule
Transfer Setting & Target Domain & Original on Target & Transfer Acc. & Drop \\
\midrule
GSM8K $\rightarrow$ SVAMP 
& Math $\rightarrow$ Math 
& 94.72 & \textbf{93.85} & 0.87 \\
HumanEval $\rightarrow$ DS-1000 
& Code $\rightarrow$ Code 
& 58.65 & \textbf{57.42} & 1.23 \\
GSM8K $\rightarrow$ HumanEval 
& Math $\rightarrow$ Code 
& 94.36 & \textbf{88.54} & 5.82 \\
HumanEval $\rightarrow$ GSM8K 
& Code $\rightarrow$ Math 
& 97.60 & \textbf{90.71} & 6.89 \\
\bottomrule
\end{tabular}
}
\caption{
Generalization study via cross-dataset transfer. 
$A \rightarrow B$ denotes that the scheduler is few-shot trained on dataset $A$ and directly evaluated on dataset $B$. 
``Original on Target'' denotes the performance of \ourmethod trained directly on the target dataset, while ``Transfer Acc.'' denotes the performance under the transfer setting.
}
\label{tab:generalization_transfer}
\end{table}

\edit{As shown in Table~\ref{tab:generalization_transfer}, \ourmethod exhibits strong transferability when the source and target datasets belong to the same task domain. 
For GSM8K $\rightarrow$ SVAMP, the transferred scheduler achieves $93.85\%$ accuracy, only $0.87\%$ lower than \ourmethod trained on the target dataset. 
Similarly, for HumanEval $\rightarrow$ DS-1000, the transferred scheduler obtains $57.42\%$ accuracy, with a moderate drop of $1.23\%$. 
These results suggest that the scheduling trajectory memory and learned scheduler parameters can capture reusable coordination patterns within the same type of reasoning task.}

\edit{Cross-domain transfer is more challenging. 
When transferring from GSM8K to HumanEval, the accuracy decreases from $94.36\%$ to $88.54\%$, and transferring from HumanEval to GSM8K decreases the accuracy from $97.60\%$ to $90.71\%$. 
This indicates that mathematical reasoning and code reasoning require different communication patterns among agents. 
Nevertheless, the transferred models still achieve competitive performance and remain stronger than static multi-agent systems, suggesting that \ourmethod learns non-trivial scheduling behaviors that can partially generalize beyond the source domain.}

\subsection{Zero-Shot Cross-Task Transfer}
\label{app:zero_shot_transfer}

The preceding experiment allows limited target-task adaptation. We further evaluate a stricter zero-shot setting in which a scheduler trained on source task $A$ is directly applied to target task $B$ without any target-task training query.

\begin{table*}[!htbp]
\centering
\small
\resizebox{0.9\textwidth}{!}{
\begin{tabular}{llccc}
\toprule
\textbf{Transfer Setting} & \textbf{Domain} & \textbf{Target-Trained} & \textbf{Zero-Shot} & \textbf{Drop} \\
\midrule
GSM8K $\rightarrow$ SVAMP & Math $\rightarrow$ Math & 94.72 & 91.38 & 3.34 \\
GSM8K $\rightarrow$ MultiArith & Math $\rightarrow$ Math & 99.05 & 96.73 & 2.32 \\
HumanEval $\rightarrow$ DS-1000 & Code $\rightarrow$ Code & 58.65 & 53.07 & 5.58 \\
\midrule
GSM8K $\rightarrow$ HumanEval & Math $\rightarrow$ Code & 94.36 & 81.92 & 12.44 \\
HumanEval $\rightarrow$ GSM8K & Code $\rightarrow$ Math & 97.60 & 84.17 & 13.43 \\
GSM8K $\rightarrow$ MMLU & Math $\rightarrow$ General & 89.85 & 80.26 & 9.59 \\
HumanEval $\rightarrow$ HotpotQA & Code $\rightarrow$ Web & 92.52 & 79.63 & 12.89 \\
\bottomrule
\end{tabular}}
\caption{Zero-shot cross-task transfer. Target-Trained denotes a scheduler trained on the target dataset; Zero-Shot directly transfers the source scheduler without target-task queries. All values are accuracy (\%).}
\label{tab:zero_shot_transfer}
\end{table*}

Table~\ref{tab:zero_shot_transfer} quantifies a clear domain gap. In-domain zero-shot transfer incurs a $2.32$--$5.58$ point drop, whereas cross-domain transfer incurs a $9.59$--$13.43$ point drop. Together with Table~\ref{tab:generalization_transfer}, these results show that the scheduler learns reusable behaviors within a task family, while a small amount of target-domain adaptation substantially closes the gap for structurally different tasks.

\section{Details of Baselines}
\edit{For fair comparison, all baselines use \texttt{gpt-oss-120b}, the same prompt template as \ourmethod, and decoding temperature \(0.1\). 
For method-specific hyperparameters, we follow the original papers whenever available and tune validation-dependent thresholds under the same training budget of \(40\)--\(80\) queries.}

\subsection{Positioning Relative to Agent Harnesses}

Agent harnesses address a broader engineering problem than communication-topology scheduling: they may provide tool integration, memory management, sandboxing, sub-agent creation, and deployment infrastructure. Representative commercial and research systems---including Claude Code, Codex, OpenHands, and MASAI---primarily expose execution or workflow abstractions. By contrast, \ourmethod is a learned scheduling module whose output is the communication graph used at each dialogue iteration. It does not replace a harness and is agnostic to its tool and sandbox implementations; instead, it can be integrated into a harness to replace a fixed or manually specified inter-agent communication pattern. This distinction also explains why harness-level systems are not treated as directly comparable topology-learning baselines in the main experiments.

\begin{table}[!htbp]
\centering
\small
\resizebox{1\columnwidth}{!}{
\begin{tabular}{lc}
\toprule
\textbf{Hyperparameter} & \textbf{Value} \\
\midrule
Backbone LLM & \texttt{gpt-oss-120b} \\
Prompt template & Same as \ourmethod \\
Decoding temperature & \(0.1\) \\
Optimization target & Graph edges and prompts \\
Graph optimizer & REINFORCE \\
Initial edge probability & \(0.10\) \\
Learning rate & \(0.40\) \\
Sampled graphs per iteration & \(20\) \\
Training budget & \(40\)--\(80\) queries \\
Stopping criterion & Budget exhausted \\
\bottomrule
\end{tabular}}
\caption{Key hyperparameter settings of GPTSwarm.}
\label{tab:gptswarm_hparams}
\end{table}

\begin{table}[!htbp]
\centering
\small
\resizebox{1\columnwidth}{!}{
\begin{tabular}{lc}
\toprule
\textbf{Hyperparameter} & \textbf{Value} \\
\midrule
Backbone LLM & \texttt{gpt-oss-120b} \\
Prompt template & Same as \ourmethod \\
Decoding temperature & \(0.1\) \\
Communication rounds & \(3\) \\
Node encoder & \texttt{all-MiniLM-L6-v2} \\
Node embedding dimension & \(384\) \\
Topology sampling times & \(10\) \\
Anchor regularization weight & \(0.01\) \\
Sparsity regularization weight & \(0.10\) \\
Training budget & \(40\)--\(80\) queries \\
\bottomrule
\end{tabular}}
\caption{Key hyperparameter settings of G-Designer.}
\label{tab:gdesigner_hparams}
\end{table}

\begin{table}[!htbp]
\centering
\small
\resizebox{1\columnwidth}{!}{
\begin{tabular}{lc}
\toprule
\textbf{Hyperparameter} & \textbf{Value} \\
\midrule
Backbone LLM & \texttt{gpt-oss-120b} \\
Prompt template & Same as \ourmethod \\
Decoding temperature & \(0.1\) \\
Search algorithm & Monte Carlo Tree Search \\
Workflow representation & Code-represented workflow \\
Optimized components & Workflow edges and node prompts \\
Workflow operator set & Original AFlow operators \\
Maximum optimization rounds & \(20\) \\
Training budget & \(40\)--\(80\) queries \\
Early stopping criterion & Validation score not improved \\
\bottomrule
\end{tabular}}
\caption{Key hyperparameter settings of AFlow.}
\label{tab:aflow_hparams}
\end{table}

\begin{table}[!htbp]
\centering
\small
\resizebox{1\columnwidth}{!}{
\begin{tabular}{lc}
\toprule
\textbf{Hyperparameter} & \textbf{Value} \\
\midrule
Backbone LLM & \texttt{gpt-oss-120b} \\
Prompt template & Same as \ourmethod \\
Decoding temperature & \(0.1\) \\
Routing granularity & Reasoning step \\
Draft model & \texttt{gpt-oss-120b} \\
Verifier model & \texttt{gpt-oss-120b} \\
Verification strategy & LLM-as-judge \\
Utility score range & \(0\) to \(9\) \\
Acceptance threshold candidates & \(3, 5, 7, 9\) \\
Acceptance threshold selection & Validation-selected \\
Fallback strategy & Regenerate rejected step \\
Training budget & \(40\)--\(80\) queries \\
\bottomrule
\end{tabular}}
\caption{Key hyperparameter settings of SpecReason.}
\label{tab:specreason_hparams}
\end{table}

\begin{table}[!htbp]
\centering
\small
\resizebox{1\columnwidth}{!}{
\begin{tabular}{lc}
\toprule
\textbf{Hyperparameter} & \textbf{Value} \\
\midrule
Backbone LLM & \texttt{gpt-oss-120b} \\
Prompt template & Same as \ourmethod \\
Decoding temperature & \(0.1\) \\
Routing granularity & Reasoning step \\
Routing signal & Logit-based confidence \\
Confidence estimator & Maximum logit confidence \\
Calibration model & Two-component Gaussian mixture model \\
External router & No \\
Routing threshold search interval & \(0.10\) \\
Routing threshold selection & Validation-selected \\
Fallback strategy & Route to high-confidence branch \\
Training budget & \(40\)--\(80\) queries \\
\bottomrule
\end{tabular}}
\caption{Key hyperparameter settings of STEER.}
\label{tab:steer_hparams}
\end{table}

\begin{table}[!htbp]
\centering
\small
\resizebox{1\columnwidth}{!}{
\begin{tabular}{lc}
\toprule
\textbf{Hyperparameter} & \textbf{Value} \\
\midrule
Backbone LLM & \texttt{gpt-oss-120b} \\
Prompt template & Same as \ourmethod \\
Decoding temperature & \(0.1\) \\
Initial topology & Fully connected graph \\
Pruning granularity & Communication edge \\
Initial edge scoring & Semantic compatibility \\
Adaptive edge scoring & Historical contribution \\
Clustering strategy & 0-extension clustering \\
Pruning schedule & Progressive pruning \\
Greedy top-k pruning & No \\
Training budget & \(40\)--\(80\) queries \\
Final topology selection & Validation-selected \\
\bottomrule
\end{tabular}}
\caption{Key hyperparameter settings of SafeSieve.}
\label{tab:safesieve_hparams}
\end{table}

\begin{table}[!htbp]
\centering
\small
\resizebox{1\columnwidth}{!}{
\begin{tabular}{lc}
\toprule
\textbf{Hyperparameter} & \textbf{Value} \\
\midrule
Backbone LLM & \texttt{gpt-oss-120b} \\
Prompt template & Same as \ourmethod \\
Decoding temperature & \(0.1\) \\
Topology generation paradigm & Autoregressive graph generation \\
Graph initialization & Empty graph \\
Node generation strategy & Role selection from agent pool \\
Edge generation strategy & Sequential edge prediction \\
Node encoder & \texttt{all-MiniLM-L6-v2} \\
Node embedding dimension & \(384\) \\
Communication rounds & \(3\) \\
Training strategy & Two-stage curriculum learning \\
Training objective & Negative log-likelihood \\
Node-edge loss weight & \(0.20\) \\
Training budget & \(40\)--\(80\) queries \\
Final aggregation & Summarizer agent \\
\bottomrule
\end{tabular}}
\caption{Key hyperparameter settings of ARG-Designer.}
\label{tab:argdesigner_hparams}
\end{table}

\section{Case Study}
\begin{figure}[!htbp]
    \centering
\includegraphics[width=0.85\linewidth]{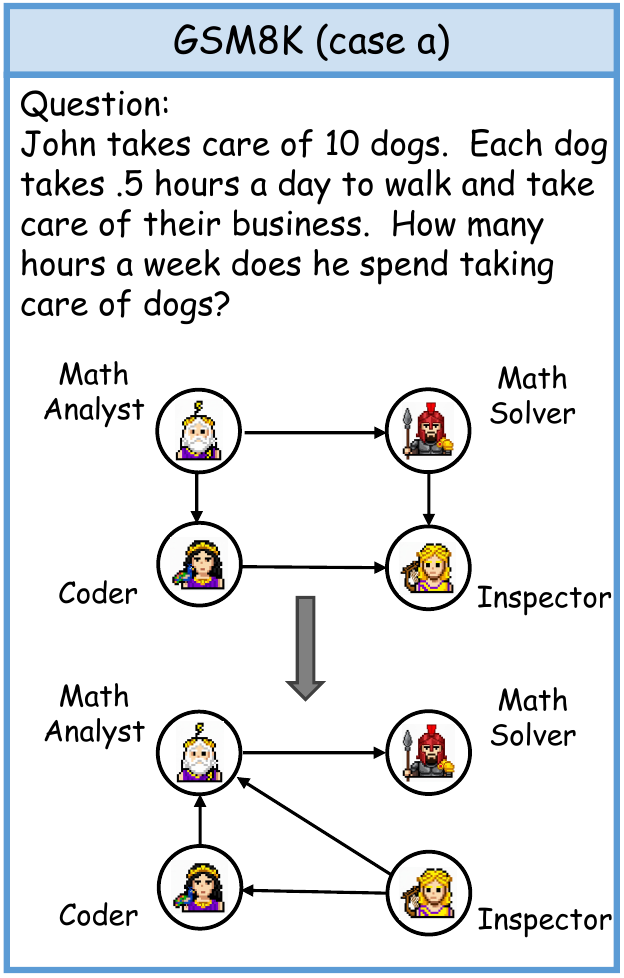}
    \caption{Case study on GSM8K.}
    \label{fig: case a}
\end{figure}

\begin{figure}[!htbp]
    \centering
\includegraphics[width=0.85\linewidth]{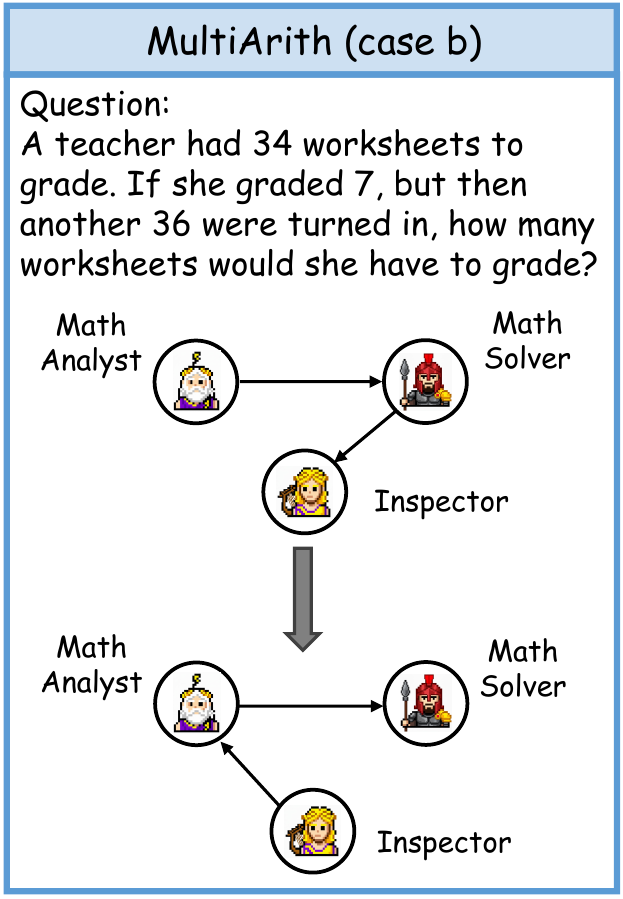}
    \caption{Case study on MultiArith.}
    \label{fig: case b}
    
\end{figure}

\begin{figure}[!htbp]
    \centering
\includegraphics[width=0.9\linewidth]{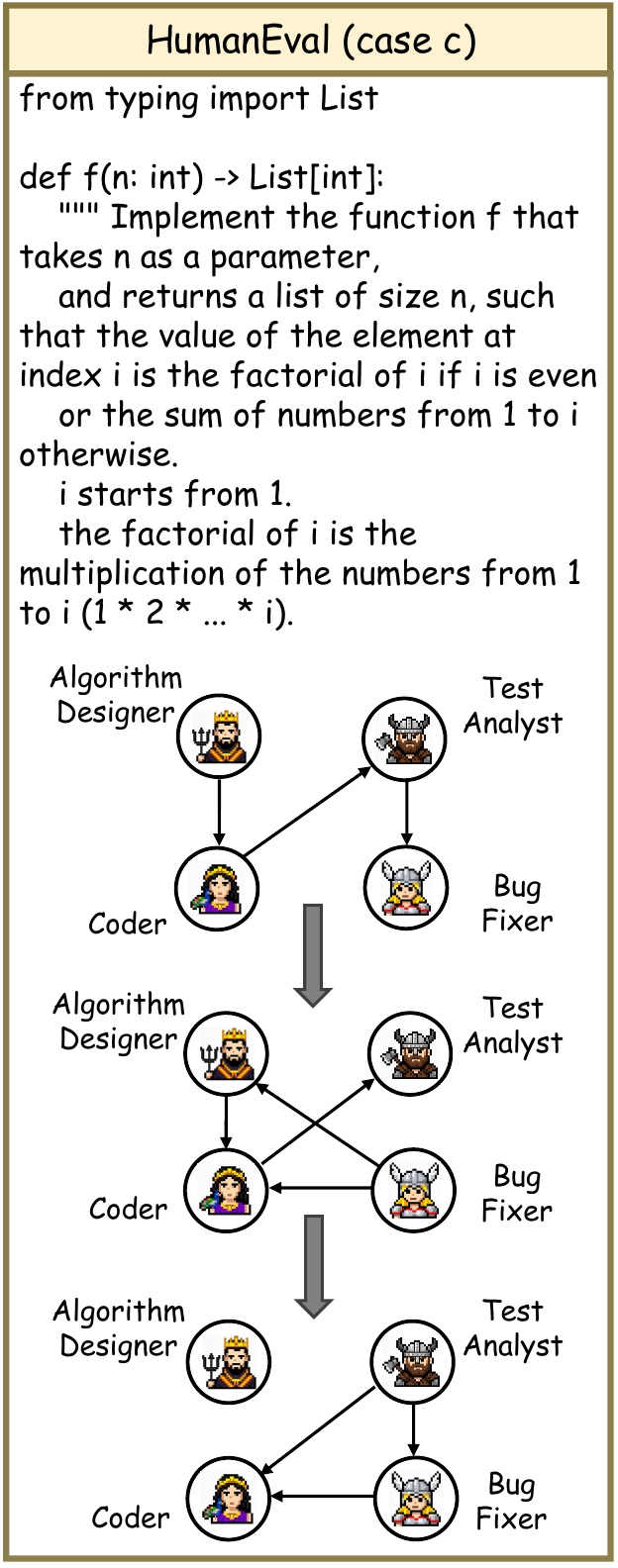}
    \caption{Case study on HumanEval.}
    \label{fig: case c}
    
\end{figure}

\begin{figure}[!htbp]
    \centering
\includegraphics[width=0.9\linewidth]{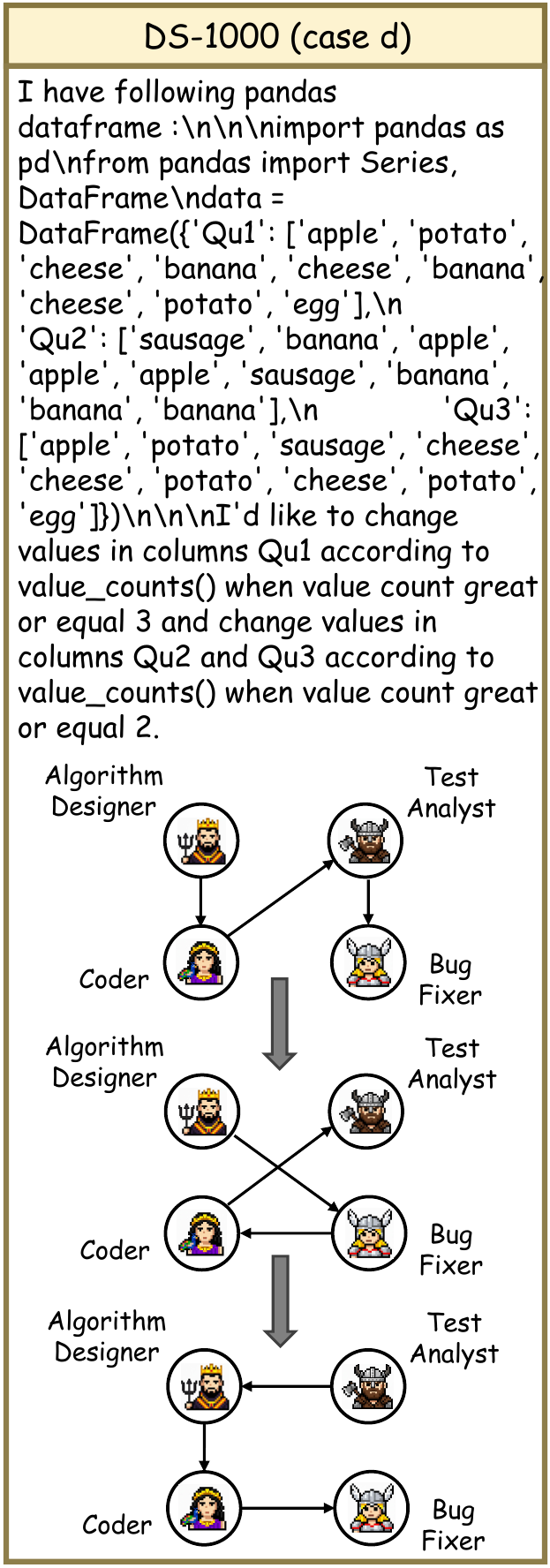}
    \caption{Case study on DS-1000.}
    \label{fig: case d}
    
\end{figure}

\begin{figure}[!htbp]
    \centering
\includegraphics[width=0.9\linewidth]{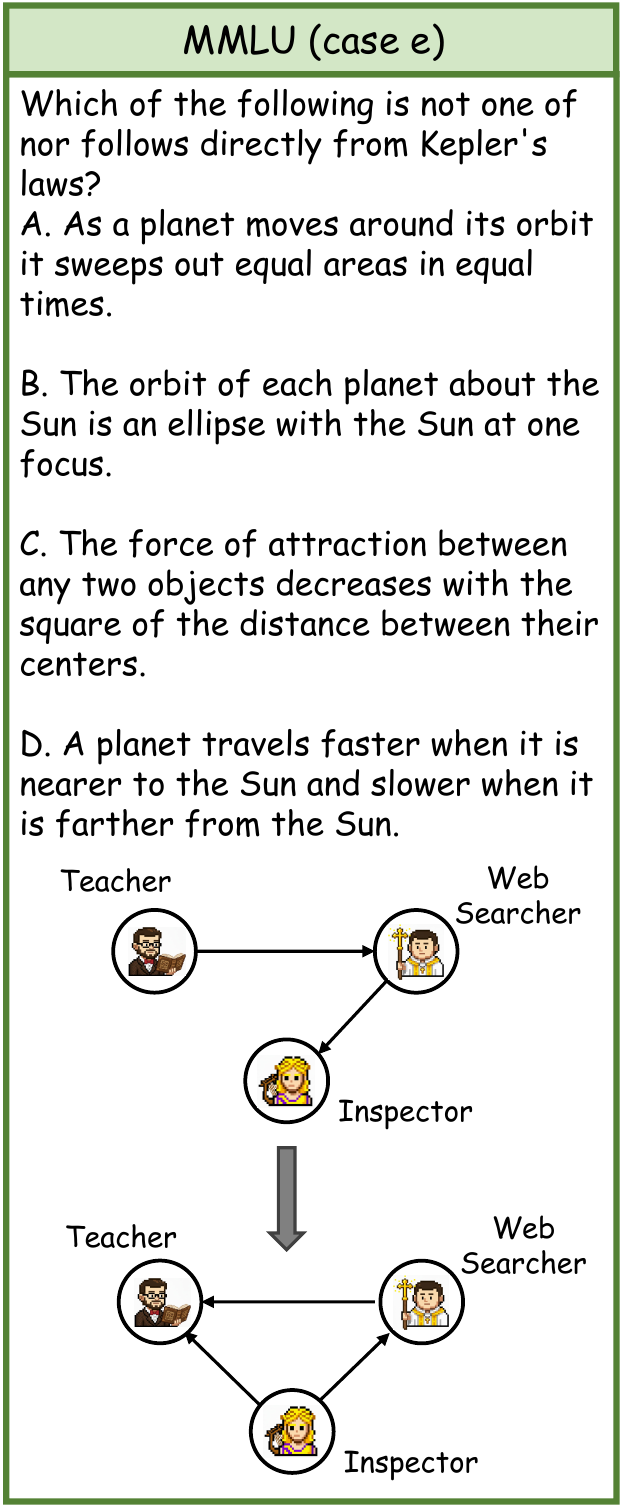}
    \caption{Case study on MMLU.}
    \label{fig: case e}
    
\end{figure}

We showcase some scheduling trajectories in this section--see Figure~\ref{fig: case a}--\ref{fig: case e}.

\section{More Robustness Analyses}
\label{app:robustness_compromised_agents}

\edit{We further evaluate the robustness of \ourmethod under different fractions of compromised agents. 
We consistently use five agents for all multi-agent systems and conduct experiments on MMLU, SVAMP, and HumanEval, which respectively cover question answering, mathematical reasoning, and code generation. 
For each system, Attack $(k/5)$ denotes that $k$ out of five agents are injected with an adversarial system prompt that instructs them to produce plausible but incorrect answers. 
We compare \ourmethod with Random Graph, a static multi-agent system with randomly constructed communication topology, and SafeSieve, a stronger scheduling-based baseline.}

\begin{table*}[!htbp]
\centering
\small
\setlength{\tabcolsep}{4pt}
\renewcommand{\arraystretch}{1.08}
\resizebox{0.85\textwidth}{!}{
\begin{tabular}{llcccc}
\toprule
Method & Attack Ratio & MMLU & SVAMP & HumanEval & Avg. \\
\midrule

\multirow{4}{*}{Random Graph}
& Clean $(0/5)$  & 83.20 & 87.10 & 84.14 & 84.81 \\
& Attack $(1/5)$ & 76.48 {\scriptsize ($\downarrow$6.72)} & 79.72 {\scriptsize ($\downarrow$7.38)} & 77.73 {\scriptsize ($\downarrow$6.41)} & 77.98 {\scriptsize ($\downarrow$6.83)} \\
& Attack $(2/5)$ & 72.36 {\scriptsize ($\downarrow$10.84)} & 75.64 {\scriptsize ($\downarrow$11.46)} & 73.97 {\scriptsize ($\downarrow$10.17)} & 73.99 {\scriptsize ($\downarrow$10.82)} \\
& Attack $(3/5)$ & 67.83 {\scriptsize ($\downarrow$15.37)} & 70.82 {\scriptsize ($\downarrow$16.28)} & 68.52 {\scriptsize ($\downarrow$15.62)} & 69.06 {\scriptsize ($\downarrow$15.75)} \\
\midrule

\multirow{4}{*}{SafeSieve}
& Clean $(0/5)$  & 84.65 & 90.10 & 90.15 & 88.30 \\
& Attack $(1/5)$ & 80.93 {\scriptsize ($\downarrow$3.72)} & 85.15 {\scriptsize ($\downarrow$4.95)} & 84.31 {\scriptsize ($\downarrow$5.84)} & 83.46 {\scriptsize ($\downarrow$4.84)} \\
& Attack $(2/5)$ & 78.47 {\scriptsize ($\downarrow$6.18)} & 82.03 {\scriptsize ($\downarrow$8.07)} & 81.03 {\scriptsize ($\downarrow$9.12)} & 80.51 {\scriptsize ($\downarrow$7.79)} \\
& Attack $(3/5)$ & 75.01 {\scriptsize ($\downarrow$9.64)} & 78.58 {\scriptsize ($\downarrow$11.52)} & 76.42 {\scriptsize ($\downarrow$13.73)} & 76.67 {\scriptsize ($\downarrow$11.63)} \\
\midrule

\multirow{4}{*}{\ourmethod}
& Clean $(0/5)$  & \textbf{89.20} & \textbf{93.70} & \textbf{93.20} & \textbf{92.03} \\
& Attack $(1/5)$ & \textbf{88.57} {\scriptsize ($\downarrow$0.63)} & \textbf{93.49} {\scriptsize ($\downarrow$0.21)} & \textbf{92.23} {\scriptsize ($\downarrow$0.97)} & \textbf{91.43} {\scriptsize ($\downarrow$0.60)} \\
& Attack $(2/5)$ & \textbf{87.34} {\scriptsize ($\downarrow$1.86)} & \textbf{92.07} {\scriptsize ($\downarrow$1.63)} & \textbf{90.31} {\scriptsize ($\downarrow$2.89)} & \textbf{89.91} {\scriptsize ($\downarrow$2.12)} \\
& Attack $(3/5)$ & \textbf{85.08} {\scriptsize ($\downarrow$4.12)} & \textbf{89.96} {\scriptsize ($\downarrow$3.74)} & \textbf{86.99} {\scriptsize ($\downarrow$6.21)} & \textbf{87.34} {\scriptsize ($\downarrow$4.69)} \\
\bottomrule
\end{tabular}
}
\caption{Robustness analyses under different fractions of compromised agents. We consistently use five agents for all multi-agent systems. We report the accuracy (\%) on MMLU, SVAMP, and HumanEval, as well as the average performance. Attack $(k/5)$ denotes that $k$ out of five agents are injected with an adversarial system prompt. The values in parentheses indicate the absolute performance drop compared with the corresponding clean setting.}
\label{tab:robust_attack_ratio_drop}
\end{table*}

\edit{As shown in Table~\ref{tab:robust_attack_ratio_drop}, \ourmethod consistently suffers the smallest performance degradation under all attack ratios. 
When only one agent is compromised, \ourmethod only drops by $0.60\%$ on average, while Random Graph and SafeSieve drop by $6.83\%$ and $4.84\%$, respectively. 
As the attack ratio increases to Attack $(2/5)$, the average drop of \ourmethod remains limited to $2.12\%$, which is substantially smaller than the drops of Random Graph ($10.82\%$) and SafeSieve ($7.79\%$). 
Even under the strongest Attack $(3/5)$ setting, where more than half of the agents are compromised, \ourmethod still preserves an average accuracy of $87.34\%$, with a drop of only $4.69\%$. 
In contrast, Random Graph and SafeSieve suffer much larger average drops of $15.75\%$ and $11.63\%$, respectively.}

\edit{These results demonstrate that \ourmethod is more robust to adversarial agents than both static and scheduling-based baselines. 
The static topology in Random Graph makes the system vulnerable to corrupted information propagation, since compromised agents can continuously influence the same communication paths. 
SafeSieve improves robustness by introducing scheduling mechanisms, but still suffers notable degradation as the number of compromised agents increases. 
By contrast, \ourmethod dynamically adjusts the communication topology across reasoning steps, which helps reduce the influence of unreliable agents and preserve useful information flow among normal agents. 
This confirms that the learned spatio-temporal scheduling mechanism is effective not only for improving accuracy under clean settings, but also for maintaining stable reasoning performance under adversarial perturbations.}

\section{More Scalability Analyses}
\label{app:more_scalability}

\edit{We provide more scalability analyses from three complementary perspectives: the number of agents, the number of reasoning iterations, and the number of training queries. 
The corresponding results are reported in Tables~\ref{tab:scalability_mmlu}--\ref{tab:scalability_svamp}, Tables~\ref{tab:iteration_mmlu}--\ref{tab:iteration_svamp}, and Table~\ref{tab:training_queries}, respectively.}

\paragraph{Scalability with respect to the number of agents.}
\edit{Tables~\ref{tab:scalability_mmlu}--\ref{tab:scalability_svamp} compare different methods under increasing numbers of agents. 
Across MMLU, HumanEval, and SVAMP, \ourmethod consistently achieves the best accuracy under all agent scales. 
More importantly, \ourmethod exhibits a favorable accuracy--efficiency trade-off when the number of agents increases. 
For example, on MMLU, increasing the agent number from 5 to 40 improves the accuracy of \ourmethod from $89.2\%$ to $94.1\%$, while the time cost remains lower than most optimization-based baselines at large scales. 
Similarly, on SVAMP, \ourmethod achieves the highest accuracy with the lowest time and token consumption across all agent numbers. 
These results indicate that the scheduler can effectively exploit additional agents while avoiding excessive communication overhead.}

\paragraph{Scalability with respect to the number of reasoning iterations.}
\edit{Tables~\ref{tab:iteration_mmlu}--\ref{tab:iteration_svamp} further evaluate the scalability of different methods under varying numbers of reasoning iterations. 
Compared with baselines, \ourmethod maintains consistently higher accuracy as the number of iterations increases. 
Meanwhile, the additional cost introduced by more iterations grows moderately, showing that \ourmethod does not rely on exhaustive multi-round communication. 
This is particularly important for multi-agent reasoning, since naively increasing the number of iterations can quickly amplify both latency and token consumption. 
The results suggest that \ourmethod can obtain stable gains from additional reasoning steps while preserving computational efficiency.}

\paragraph{Scalability with respect to the number of training queries.}
\edit{Table~\ref{tab:training_queries} studies how the performance of \ourmethod changes with different numbers of training queries. 
The results show a clear and consistent improvement as the number of training queries increases from 10 to 80. 
With only 10 training queries, \ourmethod already outperforms strong baselines such as LLM-Debate and G-Designer on average. 
As more training queries are used, the performance further improves across diverse benchmarks, including general question answering, mathematical reasoning, coding, web reasoning, and medical reasoning. 
This indicates that the scheduler can effectively benefit from additional few-shot training trajectories and gradually learn more generalizable communication patterns.}

\edit{Overall, these scalability analyses demonstrate that \ourmethod scales well along three important axes: agent size, reasoning depth, and training data size. 
It consistently improves accuracy while maintaining relatively low time and token costs, suggesting that the proposed scheduler provides an effective and efficient mechanism for coordinating multi-agent reasoning systems at larger scales.}
\begin{table*}[!htbp]
\centering

\resizebox{0.65\textwidth}{!}{
\begin{tabular}{llcccc}
\toprule
\textbf{Method} & \textbf{Metric} & \textbf{5 Agents} & \textbf{10 Agents} & \textbf{20 Agents} & \textbf{40 Agents} \\
\midrule
Random Graph & Accuracy (\%) & 82.7 & 83.3 & 83.6 & 83.8 \\
             & Time (minutes) & 14.2 & 31.4 & 59.7 & 118.5 \\
             & \#Tokens & 431,064 & 1,317,148 & 4,245,925 & 13,184,762 \\
\midrule
GPTSwarm & Accuracy (\%) & 83.6 & 84.2 & 84.8 & 85.1 \\
         & Time (minutes) & 67.8 & 201.8 & 422.5 & 825.0 \\
         & \#Tokens & 1,672,201 & 5,103,328 & 14,054,825 & 39,487,316 \\
\midrule
G-Designer & Accuracy (\%) & 84.1 & 85.2 & 87.3 & 88.1 \\
           & Time (minutes) & 22.4 & 41.8 & 75.23 & 138.6 \\
           & \#Tokens & 1,422,309 & 2,887,532 & 6,130,221 & 12,873,904 \\
\midrule
AFlow & Accuracy (\%) & 83.0 & 84.5 & 85.1 & 85.5 \\
      & Time (minutes) & 36.7 & 77.2 & 135.4 & 248.0 \\
      & \#Tokens & 1,851,925 & 3,485,925 & 7,292,296 & 15,176,842 \\
\midrule
STEER & Accuracy (\%) & 84.0 & 86.2 & 87.8 & 88.7 \\
      & Time (minutes) & 48.2 & 105.6 & 233.2 & 492.0 \\
      & \#Tokens & 2,025,125 & 4,256,252 & 9,026,290 & 19,367,814 \\
\midrule
\ourmethod & Accuracy (\%) & \textbf{89.2} & \textbf{91.6} & \textbf{93.3} & \textbf{94.1} \\
       & Time (minutes) & 15.2 & \textbf{28.4} & \textbf{51.6} & \textbf{94.8} \\
       & \#Tokens & 1,162,296 & 1,825,926 & \textbf{3,326,037} & \textbf{5,942,183} \\
\bottomrule
\end{tabular}}
\caption{Scalability comparison on MMLU under different numbers of agents.}
\label{tab:scalability_mmlu}
\end{table*}

\begin{table*}[!htbp]
\centering
\resizebox{0.65\textwidth}{!}{
\begin{tabular}{llcccc}
\toprule
\textbf{Method} & \textbf{Metric} & \textbf{5 Agents} & \textbf{10 Agents} & \textbf{20 Agents} & \textbf{40 Agents} \\
\midrule
Random Graph & Accuracy (\%) & 83.6 & 84.1 & 84.6 & 84.9 \\
             & Time (minutes) & \textbf{3.8} & \textbf{7.5} & \textbf{15.2} & \textbf{31.4} \\
             & \#Tokens & 158,426 & 316,917 & 641,382 & 1,296,745 \\
\midrule
GPTSwarm & Accuracy (\%) & 86.5 & 87.1 & 87.8 & 88.2 \\
         & Time (minutes) & 8.9 & 17.6 & 35.9 & 73.8 \\
         & \#Tokens & 189,742 & 381,653 & 776,438 & 1,592,817 \\
\midrule
G-Designer & Accuracy (\%) & 88.3 & 89.4 & 90.2 & 90.7 \\
           & Time (minutes) & 5.1 & 10.4 & 21.5 & 44.7 \\
           & \#Tokens & 176,385 & 352,914 & 719,682 & 1,463,259 \\
\midrule
AFlow & Accuracy (\%) & 88.4 & 89.3 & 89.9 & 90.3 \\
      & Time (minutes) & 6.4 & 13.1 & 27.2 & 56.9 \\
      & \#Tokens & 154,628 & 309,847 & 631,295 & 1,286,374 \\
\midrule
STEER & Accuracy (\%) & 88.1 & 89.0 & 89.8 & 90.4 \\
      & Time (minutes) & 6.8 & 13.9 & 29.6 & 62.7 \\
      & \#Tokens & 145,736 & 291,284 & 598,417 & 1,239,506 \\
\midrule
\ourmethod & Accuracy (\%) & \textbf{93.2} & \textbf{94.4} & \textbf{95.0} & \textbf{95.4} \\
       & Time (minutes) & 4.2 & 8.3 & 16.9 & 34.8 \\
       & \#Tokens & \textbf{138,945} & \textbf{281,376} & \textbf{574,829} & \textbf{1,176,942} \\
\bottomrule
\end{tabular}}

\caption{Scalability comparison on HumanEval under different numbers of agents.}
\label{tab:scalability_humaneval}
\end{table*}

\begin{table*}[!htbp]
\centering

\resizebox{0.65\textwidth}{!}{
\begin{tabular}{llcccc}
\toprule
\textbf{Method} & \textbf{Metric} & \textbf{5 Agents} & \textbf{10 Agents} & \textbf{20 Agents} & \textbf{40 Agents} \\
\midrule
Random Graph & Accuracy (\%) & 86.6 & 87.1 & 87.5 & 87.8 \\
             & Time (minutes) & 10.6 & 21.4 & 43.8 & 89.7 \\
             & \#Tokens & 648,372 & 1,294,816 & 2,631,457 & 5,382,904 \\
\midrule
GPTSwarm & Accuracy (\%) & 86.7 & 87.3 & 87.8 & 88.1 \\
         & Time (minutes) & 25.8 & 52.6 & 108.4 & 223.7 \\
         & \#Tokens & 1,006,481 & 2,018,736 & 4,126,592 & 8,482,317 \\
\midrule
G-Designer & Accuracy (\%) & 88.5 & 89.5 & 90.2 & 90.7 \\
           & Time (minutes) & 12.2 & 24.9 & 51.3 & 106.8 \\
           & \#Tokens & 552,946 & 1,108,375 & 2,264,831 & 4,682,519 \\
\midrule
AFlow & Accuracy (\%) & 85.9 & 86.5 & 87.0 & 87.4 \\
      & Time (minutes) & 21.7 & 44.3 & 91.5 & 189.6 \\
      & \#Tokens & 904,218 & 1,815,604 & 3,714,925 & 7,658,341 \\
\midrule
STEER & Accuracy (\%) & 86.8 & 87.4 & 88.1 & 88.6 \\
      & Time (minutes) & 16.3 & 33.5 & 70.1 & 146.8 \\
      & \#Tokens & 601,739 & 1,207,482 & 2,493,816 & 5,182,647 \\
\midrule
\ourmethod & Accuracy (\%) & \textbf{93.7} & \textbf{94.7} & \textbf{95.3} & \textbf{95.8} \\
       & Time (minutes) & \textbf{9.3} & \textbf{18.8} & \textbf{38.7} & \textbf{80.4} \\
       & \#Tokens & \textbf{436,215} & \textbf{872,946} & \textbf{1,785,324} & \textbf{3,692,781} \\
\bottomrule
\end{tabular}}
\caption{Scalability comparison on SVAMP under different numbers of agents.}
\label{tab:scalability_svamp}
\end{table*}

\begin{table*}[!htbp]
\centering
\small
\resizebox{0.65\textwidth}{!}{
\begin{tabular}{llccc}
\toprule
\textbf{Method} & \textbf{Metric} & \textbf{3 Iterations} & \textbf{6 Iterations} & \textbf{12 Iterations} \\
\midrule
Random Graph & Accuracy (\%) & 82.7 & 83.1 & 83.3 \\
             & Time (minutes) & 14.2 & 27.4 & 52.5 \\
             & \#Tokens & 431,064 & 816,266 & 1,425,295 \\
\midrule
GPTSwarm & Accuracy (\%) & 83.6 & 84.0 & 84.2 \\
         & Time (minutes) & 67.8 & 112.5 & 199.4 \\
         & \#Tokens & 1,672,201 & 3,025,162 & 5,925,102 \\
\midrule
G-Designer & Accuracy (\%) & 84.1 & 83.8 & 84.3 \\
           & Time (minutes) & 22.4 & 42.5 & 79.2 \\
           & \#Tokens & 1,422,309 & 3,015,285 & 5,723,293 \\
\midrule
AFlow & Accuracy (\%) & 83.0 & 83.4 & 83.6 \\
      & Time (minutes) & 36.7 & 67.2 & 122.4 \\
      & \#Tokens & 1,851,925 & 3,425,291 & 6,902,946 \\
\midrule
STEER & Accuracy (\%) & 84.0 & 84.3 & 84.5 \\
      & Time (minutes) & 48.2 & 87.2 & 144.5 \\
      & \#Tokens & 2,025,125 & 3,821,229 & 7,402,409 \\
\midrule
\ourmethod & Accuracy (\%) & \textbf{89.2} & \textbf{89.4} & \textbf{89.6} \\
       & Time (minutes) & 15.2 & \textbf{26.4} & \textbf{44.2} \\
       & \#Tokens & 1,162,296 & \textbf{1,823,394} & \textbf{2,901,034} \\
\bottomrule
\end{tabular}}
\caption{Scalability comparison on MMLU under different numbers of iterations.}
\label{tab:iteration_mmlu}
\end{table*}

\begin{table*}[!htbp]
\centering
\resizebox{0.65\textwidth}{!}{
\begin{tabular}{llccc}
\toprule
\textbf{Method} & \textbf{Metric} & \textbf{3 Iterations} & \textbf{6 Iterations} & \textbf{12 Iterations} \\
\midrule
Random Graph & Accuracy (\%) & 84.1 & 84.5 & 84.8 \\
             & Time (minutes) & \textbf{5.6} & \textbf{10.8} & \textbf{20.9} \\
             & \#Tokens & 246,382 & 467,915 & 834,276 \\
\midrule
GPTSwarm & Accuracy (\%) & 87.1 & 87.5 & 87.8 \\
         & Time (minutes) & 18.4 & 32.7 & 60.8 \\
         & \#Tokens & 381,624 & 704,892 & 1,326,517 \\
\midrule
G-Designer & Accuracy (\%) & 89.4 & 89.7 & 90.1 \\
           & Time (minutes) & 10.6 & 19.8 & 37.4 \\
           & \#Tokens & 352,768 & 671,435 & 1,238,692 \\
\midrule
AFlow & Accuracy (\%) & 89.3 & 89.6 & 89.9 \\
      & Time (minutes) & 13.2 & 24.5 & 45.7 \\
      & \#Tokens & 309,846 & 587,293 & 1,087,641 \\
\midrule
STEER & Accuracy (\%) & 89.0 & 89.4 & 89.8 \\
      & Time (minutes) & 14.7 & 26.9 & 49.6 \\
      & \#Tokens & 291,573 & 548,926 & 1,015,384 \\
\midrule
\ourmethod & Accuracy (\%) & \textbf{94.4} & \textbf{94.7} & \textbf{95.0} \\
       & Time (minutes) & 8.9 & 15.6 & 28.3 \\
       & \#Tokens & \textbf{281,394} & \textbf{492,618} & 857,936 \\
\bottomrule
\end{tabular}}
\caption{Scalability comparison on HumanEval under different numbers of iterations.}
\label{tab:iteration_humaneval}
\end{table*}

\begin{table*}[!htbp]
\centering
\resizebox{0.65\textwidth}{!}{
\begin{tabular}{llccc}
\toprule
\textbf{Method} & \textbf{Metric} & \textbf{3 Iterations} & \textbf{6 Iterations} & \textbf{12 Iterations} \\
\midrule
Random Graph & Accuracy (\%) & 86.6 & 87.0 & 87.2 \\
             & Time (minutes) & 10.6 & 20.5 & 39.2 \\
             & \#Tokens & 648,372 & 1,227,762 & 2,143,815 \\
\midrule
GPTSwarm & Accuracy (\%) & 86.7 & 87.1 & 87.3 \\
         & Time (minutes) & 25.8 & 42.8 & 75.9 \\
         & \#Tokens & 1,006,481 & 1,820,815 & 3,566,259 \\
\midrule
G-Designer & Accuracy (\%) & 88.5 & 88.2 & 88.7 \\
           & Time (minutes) & 12.2 & 23.1 & 43.1 \\
           & \#Tokens & 552,946 & 1,172,242 & 2,225,024 \\
\midrule
AFlow & Accuracy (\%) & 85.9 & 86.3 & 86.5 \\
      & Time (minutes) & 21.7 & 39.7 & 72.4 \\
      & \#Tokens & 904,218 & 1,672,427 & 3,370,422 \\
\midrule
STEER & Accuracy (\%) & 86.8 & 87.1 & 87.3 \\
      & Time (minutes) & 16.3 & 29.5 & 48.9 \\
      & \#Tokens & 601,739 & 1,135,427 & 2,199,528 \\
\midrule
\ourmethod & Accuracy (\%) & \textbf{93.7} & \textbf{93.9} & \textbf{94.1} \\
       & Time (minutes) & \textbf{9.3} & \textbf{16.2} & \textbf{27.0} \\
       & \#Tokens & \textbf{436,215} & \textbf{684,328} & \textbf{1,088,771} \\
\bottomrule
\end{tabular}}
\caption{Scalability comparison on SVAMP under different numbers of iterations.}
\label{tab:iteration_svamp}
\end{table*}

\begin{table*}[!htbp]
\centering

\resizebox{\textwidth}{!}{
\begin{tabular}{lcccccccccc}
\toprule
\textbf{Methods} & \textbf{MMLU} & \textbf{GSM8K} & \textbf{MultiArith} & \textbf{SVAMP} & \textbf{AQuA} & \textbf{HumanEval} & \textbf{DS-1000} & \textbf{HotpotQA} & \textbf{DDXPlus} & \textbf{Avg.} \\
\midrule
LLM-Debate & 83.28 & 89.95 & 95.45 & 88.30 & 73.52 & 85.24 & 49.26 & 84.55 & 71.33 & 80.10 \\
G-Designer & 84.63 & 93.35 & 97.60 & 89.52 & 76.30 & 89.40 & 54.82 & 87.66 & 76.42 & 83.30 \\
\midrule
\ourmethod (10 Queries) & 85.35 & 93.85 & 97.45 & 89.85 & 77.60 & 89.70 & 55.05 & 88.10 & 76.95 & 83.77 \\
\ourmethod (20 Queries) & 87.10 & 95.20 & 98.20 & 91.35 & 81.10 & 91.55 & 56.35 & 89.75 & 79.15 & 85.53 \\
\ourmethod (40 Queries) & 89.85 & 96.85 & 99.05 & 93.70 & 84.90 & 93.30 & 57.65 & 91.55 & 81.40 & 87.58 \\
\ourmethod (60 Queries) & 90.10 & 97.60 & 99.12 & 94.72 & 86.10 & 94.10 & 58.30 & 92.20 & 82.25 & 88.28 \\
\ourmethod (80 Queries) & \textbf{90.25} & \textbf{97.75} & \textbf{99.18} & \textbf{94.90} & \textbf{86.56} & \textbf{94.36} & \textbf{58.65} & \textbf{92.52} & \textbf{82.50} & \textbf{88.52} \\
\bottomrule
\end{tabular}}
\caption{Performance comparison under different numbers of training queries.}
\label{tab:training_queries}
\end{table*}

\section{Deployment Cost Analysis}
\label{app:deployment_cost}

We report the standalone cost of the scheduler to complement the end-to-end time and token measurements in the main experiments. The scheduler contains 2.8M parameters, comprising a 1.2M-parameter graph encoder and a 1.6M-parameter Flow-Matching network.

\begin{table}[!htbp]
\centering
\small
\resizebox{\columnwidth}{!}{
\begin{tabular}{lrr}
\toprule
\textbf{Component} & \textbf{Time (ms)} & \textbf{Percentage} \\
\midrule
LLM inference (4 agents) & 8,500 & 93.9\% \\
Entropy computation & 45 & 0.5\% \\
Flow-Matching scheduling & 65 & 0.7\% \\
Cache retrieval & 12 & 0.1\% \\
Topology parsing and routing & 8 & 0.1\% \\
Inter-agent message passing & 420 & 4.6\% \\
\midrule
Total & 9,050 & 100\% \\
\bottomrule
\end{tabular}}
\caption{Inference-latency breakdown per dialogue iteration with four agents.}
\label{tab:latency_breakdown}
\end{table}

Table~\ref{tab:latency_breakdown} shows that entropy computation and Flow-Matching scheduling account for only $0.5\%$ and $0.7\%$ of an iteration, respectively, compared with $93.9\%$ for LLM inference. Entropy computation takes 45 ms in total: 5 ms to extract log-probabilities, 15 ms to compute token entropy, 12 ms to aggregate PE, 8 ms to compute VE, and 5 ms for state classification. Its cost scales approximately linearly with the number of agents.

\begin{table}[!htbp]
\centering
\small
\resizebox{\columnwidth}{!}{
\begin{tabular}{lc}
\toprule
\textbf{Metric} & \textbf{Value} \\
\midrule
Scheduler parameters & 2.8M \\
Training hardware & 1$\times$ NVIDIA 3090 (24 GB) \\
Training GPU memory & 4.2 GB \\
Training time (40 queries) & $\sim$35 min \\
Training time (80 queries) & $\sim$68 min \\
LLM calls for training (80 queries) & $\sim$2,400 \\
\midrule
Model loading time & 0.8 s \\
First-query scheduling latency & 72 ms \\
Steady-state scheduling latency & 65 ms \\
Inference GPU memory & 0.3 GB \\
\bottomrule
\end{tabular}}
\caption{Training and inference cost of the \ourmethod scheduler. Scheduling latency is measured with $P=100$ integration steps.}
\label{tab:scheduler_cost}
\end{table}

\begin{table}[!htbp]
\centering
\small
\resizebox{\columnwidth}{!}{
\begin{tabular}{lccc}
\toprule
\textbf{Method} & \textbf{Training Time} & \textbf{GPU Memory} & \textbf{Parameters} \\
\midrule
G-Designer & $\sim$45 min & 3.8 GB & 2.1M \\
ARG-Designer & $\sim$55 min & 4.5 GB & 3.2M \\
SafeSieve & $\sim$50 min & 3.5 GB & 1.8M \\
\ourmethod & $\sim$68 min & 4.2 GB & 2.8M \\
\bottomrule
\end{tabular}}
\caption{Scheduler training-cost comparison under an 80-query budget.}
\label{tab:scheduler_training_comparison}
\end{table}

Although modeling the temporal dimension makes \ourmethod moderately more expensive to train than static topology optimizers, the scheduler remains lightweight relative to LLM inference: its cold start is below one second, steady-state topology generation takes 65 ms, and inference requires 0.3 GB of GPU memory.

\section{Reproducibility Analysis.}
\edit{We set a relatively low decoding temperature of 0.1 in our experiments, which leads to high overall reproducibility. To further verify this, we conduct experiments on all nine benchmarks, running each trial three times and reporting the mean and standard deviation in Table~\ref{tab:variance}. It can be seen that ST-EVO maintains high performance stability across different types of tasks, with standard deviations consistently below 0.5\% on all benchmarks.}

\begin{table}[!htbp]
\centering
\small
\resizebox{\columnwidth}{!}{
\begin{tabular}{lccccc}
\toprule
 & MMLU & GSM8K & MultiArith & SVAMP & AQuA \\
\midrule
ST-EVO & 89.82\scriptsize{$\pm$0.18} & 97.55\scriptsize{$\pm$0.21} & 99.01\scriptsize{$\pm$0.12} & 94.68\scriptsize{$\pm$0.24} & 86.48\scriptsize{$\pm$0.36} \\
\midrule
 & HumanEval & DS-1000 & HotpotQA & DDXPlus & \\
\midrule
ST-EVO & 94.29\scriptsize{$\pm$0.31} & 58.72\scriptsize{$\pm$0.42} & 92.49\scriptsize{$\pm$0.27} & 82.43\scriptsize{$\pm$0.39} & \\
\bottomrule
\end{tabular}}
\caption{Mean and standard deviation of ST-EVO across three independent runs.}
\label{tab:variance}
\end{table}

\end{document}